\documentclass[reprint,superscriptaddress,amsmath,amssymb,aps,prl,floatfix,longbibliography]{revtex4-2}
\usepackage{graphicx}
\graphicspath{{figures/}}
\usepackage{amsmath}
\usepackage{amssymb}
\usepackage{multirow} 
\usepackage{physics} 
\usepackage{bbold}
\usepackage{xcolor}
\usepackage{booktabs}

\usepackage[section]{placeins}

\graphicspath{{figures/}}

\AtBeginDocument{\let\oldcontentsline\contentsline}
\newcommand{\notoccontentsline}[4]{\oldcontentsline{}{}{}{}}
\newcommand{\droptocpage}{\addtocontents{toc}{\let\protect\contentsline\protect\notoccontentsline}}
\newcommand{\incltocpage}{\addtocontents{toc}{\let\protect\contentsline\protect\oldcontentsline}}

\DeclareMathOperator{\polylog}{polylog}

\begin{document}
\title{Tweezer-programmable 2D quantum walks in a Hubbard-regime lattice}
\author{Aaron W. Young}
\author{William J. Eckner}
\author{Nathan Schine}
\affiliation{JILA, University of Colorado and National Institute of Standards and Technology, and Department of Physics, University of Colorado, Boulder, Colorado 80309, USA}
\author{Andrew M. Childs}
\affiliation{Department of Computer Science, Institute for Advanced Computer Studies, and Joint Center for Quantum Information and Computer Science, University of Maryland, College Park, Maryland 20742, USA}
\author{Adam M. Kaufman}
\affiliation{JILA, University of Colorado and National Institute of Standards and Technology, and Department of Physics, University of Colorado, Boulder, Colorado 80309, USA}
\email[E-mail: ]{adam.kaufman@colorado.edu}

\begin{abstract} 

Quantum walks provide a framework for understanding and designing quantum algorithms that is both intuitive and universal. To leverage the computational power of these walks, it is important to be able to programmably modify the graph a walker traverses while maintaining coherence. Here, we do this by combining the fast, programmable control provided by optical tweezer arrays with the scalable, homogeneous environment of an optical lattice. Using this new combination of tools we study continuous-time quantum walks of single atoms on a 2D square lattice, and perform proof-of-principle demonstrations of spatial search using these walks. When scaled to more particles, the capabilities demonstrated here can be extended to study a variety of problems in quantum information science and quantum simulation, including the deterministic assembly of ground and excited states in Hubbard models with tunable interactions, and performing versions of spatial search in a larger graph with increased connectivity, where search by quantum walk can be more effective.

\end{abstract}

\date{\today}

\maketitle

\begin{figure*}[!t]
	\includegraphics[width=\linewidth]{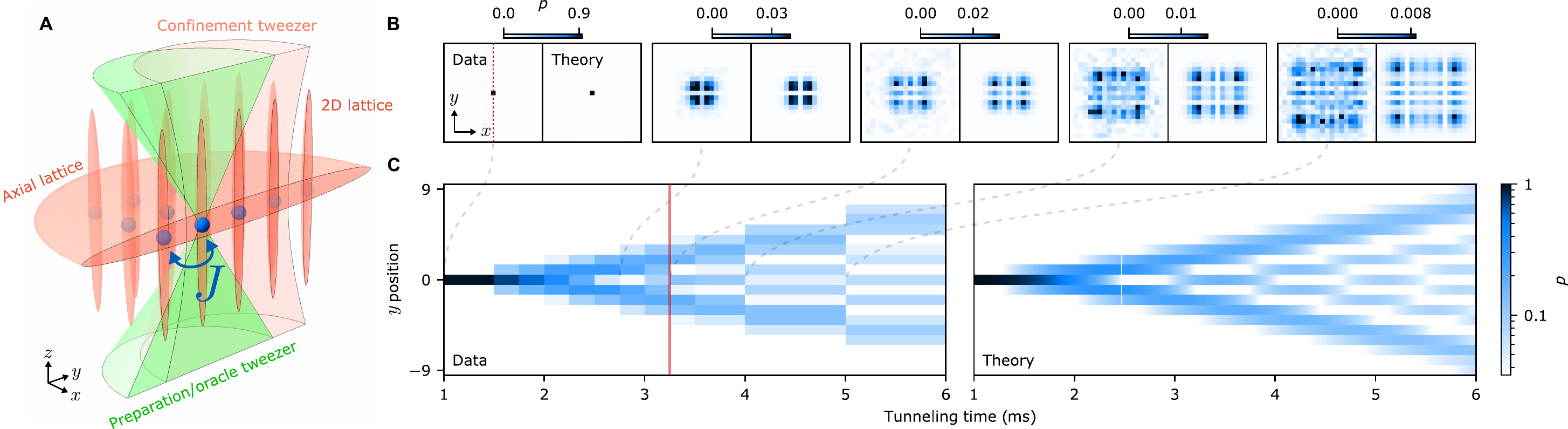}
	\caption{\textbf{Continuous-time 2D quantum walks with tweezer-implanted atoms in a lattice.} a) An array of individual $\rm ^{88}Sr$ atoms (one atom is indicated by the solid blue sphere) are loaded and cooled in optical tweezers (preparation/oracle tweezer, green), and then implanted into single sites of a 3D optical lattice composed of a 1D crossed-beam lattice aligned along gravity (axial lattice, red half-disk), and a bowtie lattice (2D lattice, red). The 2D lattice contains more than 2000 sites that are compatible with high-fidelity imaging and ground state cooling (see supplement), and is tunable to a regime where nearest-neighbor sites are coupled via a tunneling energy $J$, allowing the atoms to move through the lattice (translucent blue spheres). The preparation tweezers used to implant atoms can further be used to programmably modify the depth of individual sites in this lattice, and a large-waisted tweezer (confinement tweezer, pink) can be used to apply a tunable harmonic confining potential that spans many lattice sites. b) Atoms implanted in this lattice undergo continuous-time quantum walks in 2D, such that the probability density $p$ corresponding to their measured position (left) exhibits ballistic expansion, and wave-like interference. The behavior of these walks is in good agreement with theory using fitted values of the tunneling energy (right) up to an evolution time of 5~ms, where the atoms have coherently explored a region spanning $\sim 200$ lattice sites. Each pixel in these plots represents a single lattice site. c) Tracing out one dimension of this 2D quantum walk (in this case the $x$ axis), yields a 1D quantum walk along the remaining axis (in this case $y$, left), which is also in good agreement with theory (right) and more clearly illustrates the ballistic expansion of the wavefunction. For data to the left of the red line, multiple atoms can be implanted in different regions of the lattice for faster data collection (see supplement). Here, each point in time is averaged over 200 repetitions of the experiment. For data to the right of the red line a single atom is implanted in the center of the lattice to avoid overlapping atomic wavefunctions and averaging over inhomogeneous regions in the lattice. Here, each point up to and including 3.5~ms is averaged over 3000 repetitions. The points at 4~ms and 5~ms are averaged over 6000 and 14000 repetitions respectively.}
	\label{fig:setup}
\end{figure*}

\begin{figure*}[!t]
    \includegraphics[width=.62\linewidth]{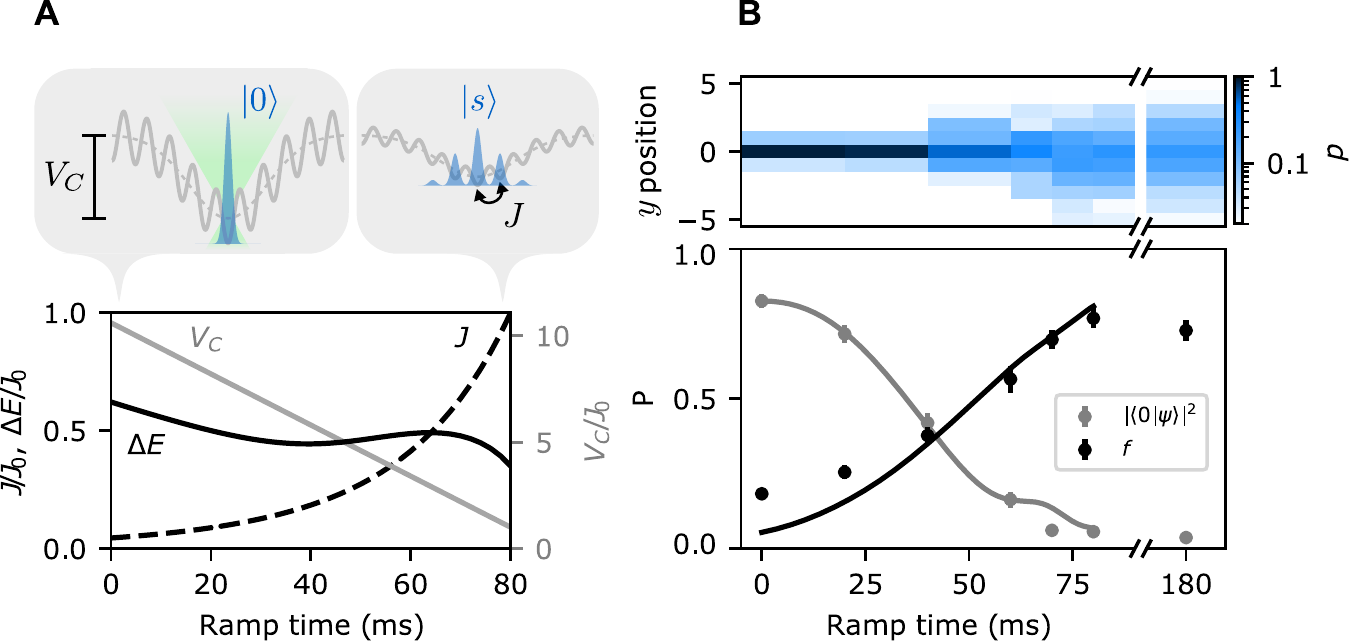}
    \caption{\textbf{Adiabatic resource state preparation.} a) An atom implanted in the lowest-energy site in a deep lattice with no tunneling (left callout), which we label as being in state $\ket{0}$, can be adiabatically connected to the ground state $\ket{s}$ of a shallow lattice with tunneling (right callout) via an adiabatic ramp of the tunneling energy, $J$, and the depth, $V_C$, of the confinement tweezer. In the callouts, the grey dashed line denotes the potential provided by the confinement tweezer, and the solid grey line the sum of the tweezer and lattice potentials. The preparation tweezer used to initially implant the atom in $\ket{0}$ is shown schematically in green. By ramping $J$ and $V_C$ together we can maintain a roughly fixed energy gap $\Delta E$ between the ground and first excited states of the system throughout the ramp. This relaxes the requirements on adiabaticity, and substantially increases the fidelity with which we are able to prepare the lattice ground state. b) As an atom evolves under this ramp its amplitude spreads over many sites in 2D. Here, the $x$ coordinate has been traced out for illustration purposes, showing the spread in the atom's $y$ position (top). During this ramp, the population on the initial site (grey points) decreases, and the overlap $f$ (see main text) between the classical probability distributions corresponding to the prepared state $\ket{\psi}$ and the expected lattice ground state $\ket{s}$ (black points) increases, eventually reaching 76.9(3.3)\%. This is in reasonable agreement with theory (solid lines) given the independently measured parameters in our ramp, and an overall scale factor to account for filtering and loss due to imperfect preparation of the atoms in their 3D motional ground states. This serves as an upper bound on the fidelity with which we can prepare the lattice ground state, but does not certify any phase coherence between components occupying different sites in the lattice. Nonetheless, this state is not observed to significantly evolve even after more than 100 tunneling times (right side of the broken axis), further suggesting that this is indeed the lattice ground state. Each data point corresponds to 500 repetitions of the experiment, except for the points at 70~ms and 180~ms, which are averaged over 1500 repetitions.}
    \label{fig:ground}
\end{figure*}

\begin{figure*}[!t]
    \includegraphics[width=.68\linewidth]{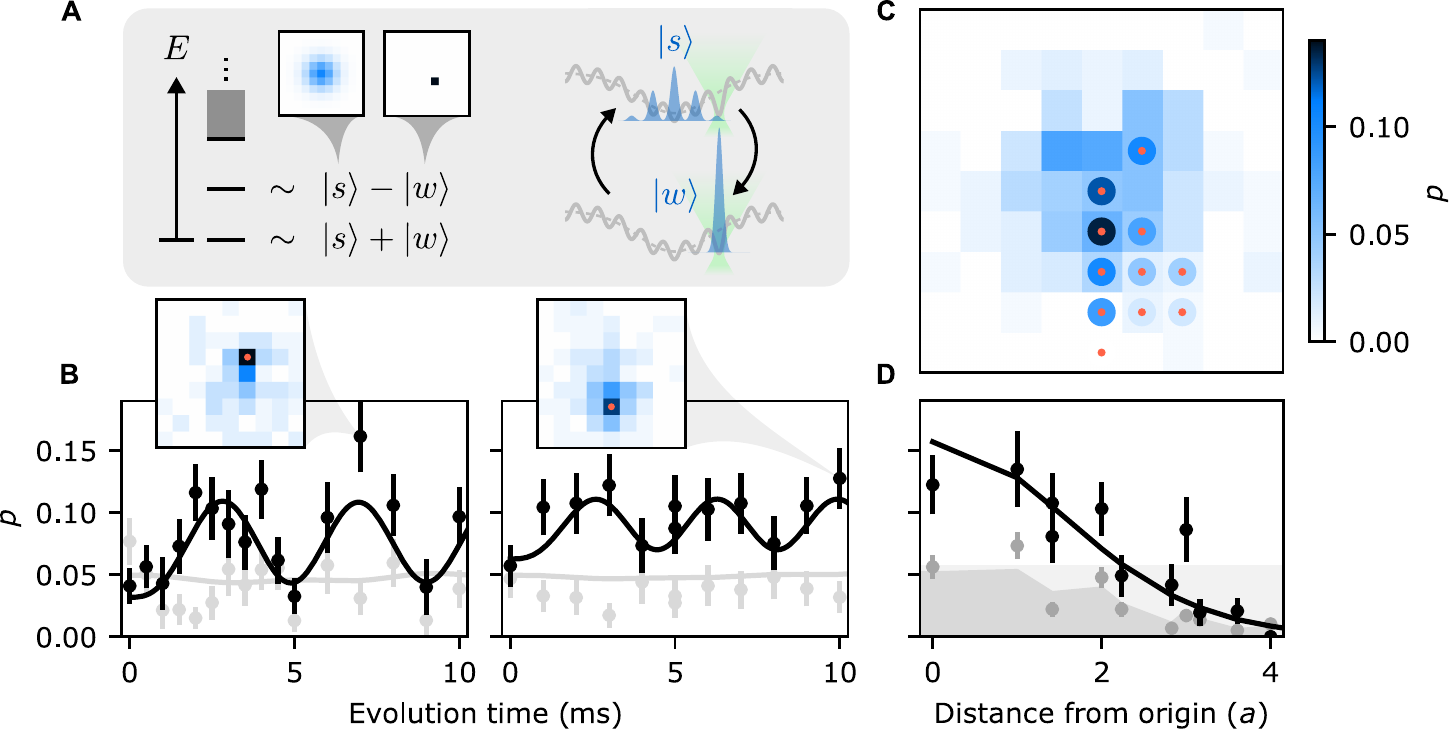}
    \caption{\textbf{Spatial search by continuous-time quantum walks.} a) When applying an oracle Hamiltonian $H_w=-V_w\ket{w}\bra{w}$ via a tweezer (green) that selects for a specific (and arbitrary) marked site $\ket{w}$ with appropriate strength, the spectrum of the system includes ground and first excited states that are approximately the even and odd superpositions of $\ket{w}$ and the lattice ground state, or resource state, $\ket{s}$. As a result, quenching to this Hamiltonian from $\ket{s}$ results in coherent oscillations between $\ket{s}$ and $\ket{w}$. b) This manifests as oscillations in the population on the marked site (black points), which we observe experimentally for a selection of oracles. These oscillations are in good agreement with theory, up to an overall offset, which uses independently characterized values of the state preparation fidelity, tunneling rate, and confinement potential (black lines). For comparison, we also plot the population in $\ket{0}$, at the center of the lattice (grey points and theory curves). Insets show the measured populations on different sites at the peak of these oscillations, where the location of the relevant oracle is marked by a red point. c) A given marked site can be found by measuring the state of the system after a half-period of these oscillations. The population on the marked site after a 2.46~ms quench is plotted for a selection of different oracle positions (each oracle is marked by a red point, with the corresponding population shown in the surrounding circle), showing a noticeable increase in amplitude relative to the amplitudes present in the prepared resource state (background image, with each pixel corresponding to a lattice site). d) To quantify the range of this search procedure we plot these amplitudes as a function of the distance of the oracle from the center of the lattice in units of the lattice spacing $a$ (black points). This is in good agreement with a theory prediction with no free parameters (black curve). At short distances, corresponding to a region spanning $\sim13$ lattice sites, the marked site can be found by looking for the highest amplitude site after the quench. At longer distances the amplitude at the origin is not significantly modified by the quench, and remains near its value in the resource state $\ket{s}$ (light grey shading), which can exceed the amplitude on the marked site. Nonetheless, in a region spanning $\sim45$ lattice sites there is a several-fold increase in the amplitude on the marked site relative to both the theoretical (dark grey shading) and measured (grey points) amplitude that was present before the quench. Color scale is shared across all parts of this figure, and all data points are averaged over 500 repetitions of the experiment, except the resource state data appearing in (c), which is averaged over 600 repetitions.
    }
    \label{fig:quench}
\end{figure*}

The ability for quantum systems to coherently explore their Hilbert space, exhibiting wave-like superposition and interference, is a key ingredient in quantum algorithms. Quantum walks are one intuitive framework for understanding the algorithmic speedups that these ingredients can provide, where states in Hilbert space are mapped to the locations of a walker on a graph~\cite{reitzner_quantum_2011}, and the walker can traverse this graph via a superposition of interfering paths. Remarkably, even in the restrictive case of real, equal-valued couplings and local, time-independent control, this simple framework is capable of universal quantum computation~\cite{childs_universal_2009, childs_universal_2013-1}, and has inspired new quantum algorithms, including those for spatial search~\cite{childs_spatial_2004-1}, graph traversal~\cite{childs_exponential_2003}, element distinctness~\cite{ambainis_quantum_2004}, and formula evaluation~\cite{farhi_quantum_2008}. When scaled to many particles, systems that realize quantum walks further allow for rich studies of quantum information and many-body physics: controlled tunneling of many non-interacting particles maps to sampling problems of interest in complexity theory~\cite{aaronson_computational_2010, muraleedharan_quantum_2019, zhong_quantum_2020}, while the combination of interactions and itinerance underlies a broad class of condensed-matter Hubbard models~\cite{lee_doping_2006, bloch_many-body_2008}. In this work, we demonstrate a new platform for realizing programmable quantum walks and lattice models that combines favorable properties of optical tweezers and optical lattices. We use this platform to perform the first demonstration of spatial search by continuous-time quantum walk with neutral atoms.

Because of their broad applicability, progressively more controllable quantum walks have been realized in a number of experimental platforms, including with photons~\cite{peruzzo_quantum_2010, sansoni_two-particle_2012, schreiber_photons_2010, schreiber_2d_2012-1, matthews_observing_2013}, nuclear magnetic resonance~\cite{ryan_experimental_2005}, matter waves~\cite{meier_atom-optics_2016}, trapped ions~\cite{schmitz_quantum_2009, zahringer_realization_2010}, and superconducting qubits~\cite{yan_strongly_2019, gong_quantum_2021-1}. Optically trapped neutral atoms are particularly amenable to realizing quantum walks~\cite{karski_quantum_2009, genske_electric_2013, preiss_strongly_2015} since they allow for high-fidelity creation and detection of individual, physically identical walkers. One approach to studying quantum walks of neutral atoms is with quantum gas microscopes, which load degenerate gases containing thousands of particles into optical lattices containing thousands of sites~\cite{bakr_quantum_2009-1, sherson_single-atom-resolved_2010}. Sophisticated techniques have been developed to ``cookie-cut" desired initial states out of these complicated many-body states~\cite{weitenberg_single-spin_2011, preiss_strongly_2015, spar_fermi-hubbard_2021-1, islam_measuring_2015}. A complementary approach is to rapidly assemble such states using optical tweezer arrays. In this case, it is possible for individual atoms to be deterministically assembled into nearly arbitrary geometries~\cite{endres_atom-by-atom_2016, barredo_atom-by-atom_2016, barredo_synthetic_2018}, and rapidly laser-cooled to their three-dimensional (3D) motional ground state~\cite{kaufman_cooling_2012, li_3d_2012, thompson_coherence_2013, norcia_microscopic_2018-1, cooper_alkaline-earth_2018, liu_molecular_2019-1, young_half-minute-scale_2020}. While pioneering studies have used tweezers to explore tunneling between up to eight sites~\cite{kaufman_two-particle_2014, murmann_two_2015-1, kaufman_entangling_2015, spar_fermi-hubbard_2021-1}, such systems are sensitive to disorder, making it difficult to realize coherent itinerance across many sites. Here we present an alternative solution that uses optical tweezers and high-fidelity laser cooling for fast, programmable implantation and control of single atoms in a Hubbard-regime optical lattice.

We use this approach to study quantum walks of individual atoms spanning hundreds of sites in a 2D lattice, and to locally control those walks to realize a spatial search algorithm. In this case, elements of the search space are represented by sites in the lattice; the location of the atom, or walker, is initialized via tweezer-implantation; and the search oracle is created with dynamically programmable tweezers superimposed on the lattice~\cite{childs_spatial_2004-1}. The techniques demonstrated here will advance studies of non-equilibrium and ground-state Hubbard physics, where fast cycle times, versatile state preparation, and site-resolved potentials can advance entanglement-measurement protocols, and implement certain sampling problems~\cite{aaronson_computational_2010, daley_measuring_2012-1, islam_measuring_2015, kaufman_quantum_2016-1, elben_renyi_2018, brydges_probing_2019-1, muraleedharan_quantum_2019}. These tools could further be extended to other types of systems that are less amenable to evaporative cooling but are powerful for many-body physics and quantum information, like molecules~\cite{liu_molecular_2019-1, anderegg_optical_2019}. 

Our experiments begin by trapping and cooling individual strontium ($\rm^{88}Sr$) atoms in optical tweezers~\cite{norcia_microscopic_2018-1}. The atoms are then transferred into one 2D layer of a 3D optical lattice (Fig.~\ref{fig:setup}a). This transfer need not be fully adiabatic with respect to on-site motional timescales, since the tightly confining optical lattice allows for even higher-fidelity optical cooling than in the tweezers, yielding a typical 3D motional ground state fraction of $100^{+0}_{-7}\%$ (see supplement). This lattice is composed of a 1D crossed-beam lattice, which is aligned along gravity, and a 2D bowtie lattice. The depths of these two lattices are independently tunable, and are reduced to $\rm26.0 E_R^{Ax}$ and $\rm5.0 E_R^{2D}$ respectively to study tunneling after optically cooling the atoms, where $\rm E_R^{Ax}$ and $\rm E_R^{2D}$ are the recoil energies of the two respective lattices (see supplement). The detuning between sites in the axial lattice due to gravity suppresses tunneling along this axis, whereas atoms tunnel freely in the 2D lattice with an average tunneling energy of $J_0/\hbar = 2\pi\times 163$~Hz, corresponding to a characteristic tunneling time of $\tau = \hbar/J_0 = 0.975$~ms (the tunneling energy differs slightly between the two axes of the lattice, see supplement).

The evolution of the atoms in this regime can be understood as a quantum walk on a graph where each site $\ket{i}$ in the lattice is represented by a node, and nodes corresponding to tunnel-coupled sites are connected by an edge. Such a graph can be represented by its adjacency matrix $A$, where $A_{ij}=0$ unless nodes $i$ and $j$ are connected, in which case $A_{ij}=1$. Given this definition, the Hamiltonian of the system is:

\begin{equation}
\label{eq:ham}
    H_{\rm Lat}= -J\sum_{i,j} A_{ij} \ket{i}\bra{j} - \sum_{i} V_{i} \ket{i}\bra{i}
\end{equation}

\noindent where $J$ is the tunneling energy, and we have further included a local energy shift $V_i$ that is present due to the finite extent of our lattice beams (see supplement), and can be programmably modified via two sets of optical tweezers (Fig.~\ref{fig:setup}a). Given the lattice uniformity, and with the tweezers fully extinguished, $\abs{V_i}\ll J$ and this term can be disregarded. In this case, an atom implanted in one site of the lattice undergoes a continuous-time quantum walk in 2D (Fig.~\ref{fig:setup}b). The evolution of this atom is in good agreement with the theoretical prediction for a flat lattice with constant $V_i$ and a distant boundary, exhibiting the formation of fringes in the probability density $p$ of the atom's measured position due to interference between the multiple paths by which the atom can arrive at a given site after its evolution. This is in contrast with the behavior of classical random walks, which exhibit diffusive expansion of a Gaussian probability density distribution. We can trace out one of the atom's two spatial coordinates (Fig.~\ref{fig:setup}c) which, given the form of $H_{\rm Lat}$ for a 2D square lattice, results in a 1D quantum walk along the remaining axis. The resulting data is in good agreement with theory, showing the expected ballistic or light-cone-like spreading of the atomic probability density~\cite{reitzner_quantum_2011}. For the latest time shown here of $t = 5.0 $~ms, the probability density continues to exhibit clear interference fringes (see supplement), suggesting that the atom has maintained phase coherence while exploring a region spanning approximately 200 lattice sites. At later times, the atom begins to sample the inhomogeneous potential resulting from the finite extent of the lattice beams, and the infinite flat lattice approximation breaks down (see supplement).

This coherent exploration of Hilbert space via continuous-time quantum walks can be harnessed in a variety of quantum algorithms, including algorithms for spatial search~\cite{childs_spatial_2004-1}. Spatial search is a specialization of the unstructured search problem with additional constraints on how the space can be explored. A continuous-time analog of Grover’s search algorithm~\cite{farhi_analog_1998} performs search by quantum walk in the limiting case of a fully connected graph. The problem becomes harder when edges are removed from the graph, which may preclude the quadratic Grover speedup depending on the graph structure. Surprisingly, quadratic speedup can persist even for much less well connected graphs. In particular, for $N$-vertex square lattice graphs the search can be performed in time $O(\sqrt{N}\polylog{N})$ in only four dimensions~\cite{childs_spatial_2004-1}, and the full, optimal~\cite{farhi_analog_1998} quadratic speedup is recovered in five or more dimensions~\cite{childs_spatial_2004-1}. Although the behavior of these single-particle walks can be captured by a classical wave equation involving $N$ coupled oscillators~\cite{grover_coupled_2002}, in the specific setting of searching a memory that is distributed in real space using a local probe~\cite{aaronson_quantum_2003-1}, quantum walks of even a single particle can provide a uniquely quantum advantage.

In these algorithms, the lack of structure in the search space suggests that a natural starting point is the uniform superposition $\ket{S} = \sum_{i=1}^N \ket{i}/\sqrt{N}$ over all standard basis states in the relevant Hilbert space. Assuming periodic boundary conditions, this resource state $\ket{S}$ is precisely the ground state of $H_{\rm Lat}$ with constant $V_i$. To approximate this resource state, we prepare the ground state $\ket{s}$ of $H_{\rm Lat}$ (Fig.~\ref{fig:ground}) in the presence of a potential $V_i$ provided by an extra confinement tweezer  (Fig.~\ref{fig:setup}a). This tweezer has a nearly-Gaussian profile with a fixed waist of $ 5.8 a $, where $a$ is the 2D lattice spacing, and tunable overall depth $V_C$. At fixed $J$, the lattice ground state is similar to $\ket{S}$ except with an approximately Gaussian envelope with a width determined by the value of $V_C$.

To prepare the state $\ket{s}$ we implant an atom in the deepest site of the combined potential generated by the lattice and the confinement tweezer, which we label as being in state $\ket{0}$. This is the ground state of the system when the lattice is deep, and $J\ll V_C$. The state $\ket{0}$ can be adiabatically connected to the ground state $\ket{s}$ in a shallow lattice via a ramp of the tunneling energy (Fig.~\ref{fig:ground}a). In practice, we perform a linear ramp of the lattice depth $V_L$ as a function of time $t$, resulting in a nonlinear ramp in the tunneling energy (see supplement). We also ramp the depth $V_C$ of the confinement tweezer to maintain an approximately constant value of the energy gap $\Delta E$ between the ground and first excited states during the ramp, which substantially relaxes the requirements on adiabaticity and improves the fidelity with which we can prepare $\ket{s}$. The observed evolution during this ramp is in reasonable agreement with theory (Fig.~\ref{fig:ground}b), where the atoms start out primarily in $\ket{0}$, and smoothly delocalize over many sites over the course of an 80~ms-long adiabatic ramp of $V_C$ and $J$. The prepared state $\ket{\psi}$ can be compared to $\ket{s}$ by computing the overlap between their populations, or the classical fidelity $f=(\sum_{i} \sqrt{p_{\psi,i} p_{s,i}})^2$, where $p_{\psi,i}$ and $p_{s,i}$ denote the populations on site $i$ in states $\ket{\psi}$ and $\ket{s}$ respectively. This constitutes an upper bound on the fidelity with which we have prepared $\ket{s}$ of 76.9(3.3)\%, but does not certify any phase coherence between the amplitudes occupying different sites. However, the prepared state is not observed to substantially evolve over more than 100 tunneling times, and the adiabatic ramp can be reversed to recover 57(5)\% of the atoms in $\ket{0}$ (excluding loss due to filtering of hot atoms, see supplement). This suggests that $\ket{s}$ has been prepared with a fidelity of 76(3)\%, in agreement with the bound set by the classical fidelity, and we can proceed with the search procedure.

The central idea in quantum-walk-based search algorithms is the presence of two competing terms in the Hamiltonian: a diffusion term $H_{\rm Lat}$ corresponding to tunneling, whose ground state is $\ket{s}$, and an oracle term ${H_w=-V_w\ket{w}\bra{w}}$, whose ground state is the marked site $\ket{w}$. $H_w$ can be applied to the system with variable drive strength $V_w$, but the choice of $\ket{w}$ is unknown to the experimenter, and is the quantity that is being searched for. Given a sufficiently connected graph and appropriate choice of $V_w$, the states $\ket{s}$ and $\ket{w}$ are similar in energy, and coupled under the full search Hamiltonian ${H=H_{\rm Lat} + H_w}$. This results in ground and first excited states $\ket{\pm}$ that are approximately the even and odd superpositions of $\ket{s}$ and $\ket{w}$, and are separated by an energy gap $\Delta E = O(1/\sqrt{N}) $, where $N$ is the number of elements in the search space (Fig.~\ref{fig:quench}a, also see supplement). As a result, quenching to this Hamiltonian from the resource state $\ket{s}$ yields coherent oscillations to $\ket{w}$ and back with a characteristic period of $O(\sqrt{N})$ that is independent of the specific choice of $\ket{w}$. Measuring the position of the walker after a half period of this oscillation identifies the marked site $\ket{w}$. This procedure can also be run backwards to prepare $\ket{s}$ from an atom implanted in a predetermined site $\ket{w'}$, avoiding any additional overhead associated with adiabatic resource state preparation (see supplement).

The choice of $V_w$ must be carefully fine-tuned to minimize the energy gap between $\ket{\pm}$ given $N$ and the connectivity of the graph~\cite{childs_spatial_2004-1}. Here, we choose $V_w = 12.55(65) J_0$, which is biased deeper than the optimal value to avoid certain sources of technical noise (see supplement). Even with optimal $V_w$, in a 2D square lattice with cyclic boundary conditions, $\ket{\pm}$ deviate from the even and odd superpositions of $\ket{s}$ and $\ket{w}$, and the scaling of $\Delta E$ with $N$ is modified, resulting in an asymptotic runtime of $\ge O(N/\polylog{N})$~\cite{childs_spatial_2004-1}. This scaling is further modified by the non-periodic boundary conditions in our experiment (see supplement). Nonetheless, upon quenching to $H$ starting in the state $\ket{s}$, we observe coherent oscillations in the population on the marked site $\ket{w}$ for a selection of different oracles (Fig.~\ref{fig:quench}b). At the peak of these oscillations, the marked site is readily identified as the highest amplitude site in the lattice. Critically, the amplitudes of these oscillations are in good agreement with theory, and limited in magnitude not by technical noise, but by the expected performance of this search procedure in a 2D square lattice (see supplement).

It is important to note that in the case of open boundary conditions, or in the presence of the confinement tweezer, the behavior of these oscillations is dependent on the specific position of $\ket{w}$ (see supplement). This position-dependent behavior sets the effective size of the search space, where at greater range reduced overlap between $\ket{w}$ and $\ket{s}$ yields oscillations with reduced amplitude. In our experiment, for an oracle a distance of $\sqrt{2}a$ away from the center of the confinement tweezer (the origin), the optimal evolution time after the quench is 2.46~ms. Performing this quench and evolution for a variety of different oracles (Fig.~\ref{fig:quench}c), we find that the amplitude on the marked site after the evolution decreases as a function of distance from the origin, in agreement with theory (Fig.~\ref{fig:quench}d). Within $2a$, corresponding to a region spanning $\sim13$ lattice sites, we can blind ourselves to the position of the oracle tweezer, and identify the marked site by looking for the most probable location of the walker after the quench. At longer range the amplitude that remains at the origin after the quench and subsequent evolution can exceed that of the marked site. However, within $\sqrt{13}a$, corresponding to a region spanning $\sim45$ lattice sites, there is still a several-fold increase in the amplitude on the marked site relative to what was present in the resource state. While not demonstrated here, in principle the effective size of the search space could be increased with constant overhead by measuring the atomic probability density as a function of evolution time after the quench, removing effects relating to variable oscillation frequencies for different oracles, as well as the large amount of amplitude that remains near the origin for distant marked sites.

In this work, we have performed a proof-of-principle demonstration of spatial search via continuous-time quantum walks of a single particle on a 2D square lattice. This is accomplished by introducing a platform that combines the programmability of optical tweezer arrays with a Hubbard-regime optical lattice that provides a clean environment for tunneling, and several thousand sites which are compatible with high-fidelity cooling, imaging, and coherent control~\cite{schine_long-lived_2021} (see supplement). Beyond studies with itinerance, these capabilities can also be used to prepare large, well-controlled ensembles of atomic qubits for quantum information, simulation, and metrology~\cite{levine_parallel_2019, graham_rydberg-mediated_2019, young_half-minute-scale_2020, ebadi_quantum_2021, scholl_quantum_2021, barnes_assembly_2021, schine_long-lived_2021, ma_universal_2021, jenkins_ytterbium_2021}.

In the specific context of spatial search, the runtime of the algorithm demonstrated in this work does not exhibit a quadratic speedup in comparison to classical search due to our use of a 2D square lattice~\cite{childs_spatial_2004-1}. A runtime of $O(\sqrt{N}\log{N})$ is achievable with a single particle in such a lattice if an additional spin-1/2 degree of freedom is used to implement a Dirac Hamiltonian~\cite{childs_spatial_2004}, or a discrete-time quantum walk~\cite{ambainis_coins_2005}. This degree of freedom can either be internal to the walker, or external, and realized using a modified optical lattice containing an array of doublets~\cite{childs_spatial_2014}. The optical clock qubit in strontium is a strong candidate for implementing this spin internally since it is well-controlled~\cite{schine_long-lived_2021}, and long-lived compared to the tunneling time~\cite{norcia_seconds-scale_2019-1, young_half-minute-scale_2020}. Moreover, it is also possible to engineer state-dependent optical potentials for this qubit~\cite{heinz_state-dependent_2020} to realize a broad class of discrete-time quantum walks~\cite{karski_quantum_2009, kitagawa_exploring_2010-1}.

While there is a setting in which such single-particle quantum walks can provide a uniquely quantum advantage~\cite{aaronson_quantum_2003-1}, this advantage can be extended to a broader class of problems by realizing these dynamics in a system whose state space scales rapidly with physical resources. This can be achieved by extending this work to multiple particles, where the state space, and thus graph size, grows exponentially in particle number (see supplement). Given the cooling performance and single-particle control demonstrated here, such experiments could be performed with tens to hundreds of atoms, where the appropriate many-body oracle is applied via tunable Rydberg-mediated interactions~\cite{schine_long-lived_2021} (see supplement). Beyond spatial search, the programmable control and assembly of large-scale itinerant systems enabled by this platform provides one route towards programmable boson sampling with many particles~\cite{aaronson_computational_2010, muraleedharan_quantum_2019}, as well as the direct assembly and characterization of Hubbard models~\cite{onari_phase_2004, ohgoe_ground-state_2012}.

% \clearpage
\droptocpage

\bibliography{references}

\section{Units}

Unless otherwise noted, all error bars and uncertainties in this article and its supplementary information are provided as one standard error of the mean.

\section{Acknowledgements}

We acknowledge fruitful discussions with E. Knill, S. Geller, and M. O. Brown. We further thank A. M. Rey, K. Kim, S. Geller, and P. M. Preiss for close readings of the manuscript.  \textbf{Funding:} This work was supported by the AFOSR (FA95501910079), ARO (W911NF1910223), the National Science Foundation Physics Frontier Center at JILA (1734006), and NIST. A. M. C. received support from the National Science Foundation (grant CCF-1813814 and QLCI grant OMA-2120757) and the Department of Energy, Office of Science, Office of Advanced Scientific Computing Research, Accelerated Research in Quantum Computing program. W. J. E. and N. S. acknowledge support from the NDSEG fellowship program, and the NRC research associateship program respectively. \textbf{Author contributions:} A. W. Y., W. J. E., N. S., and A. M. K. contributed to developing the experiments. All authors contributed to analysis of the results and preparing the manuscript. \textbf{Competing interests:} The authors declare no competing interests. \textbf{Data and materials availability:} All data in this study are available from the corresponding author upon reasonable request.

%%%%%%%%%% Merge with supplemental materials %%%%%%%%%%
\clearpage
% \widetext
% \onecolumngrid
% \appendix

%%%%%%%%%% Prefix a "S" to all equations, figures, tables and reset the counter %%%%%%%%%%
\setcounter{section}{0}
\setcounter{equation}{0}
\setcounter{figure}{0}
\setcounter{table}{0}
\setcounter{page}{1}
\makeatletter
\renewcommand{\theequation}{S\arabic{section}.\arabic{equation}}
\renewcommand{\thefigure}{S\arabic{figure}}
\renewcommand{\thetable}{S\arabic{table}}

\renewcommand{\tocname}{Supplemental Materials}
\renewcommand{\appendixname}{Supplement}

\clearpage

\tableofcontents
% \appendix
\setcounter{secnumdepth}{2}

%\begin{center}
%\textbf{\large Supplemental Materials}
%\end{center}
\incltocpage
\section{Materials and methods}

\subsection{Optical potentials and cooling}
\label{sec:potentials}

\begin{figure*}[!t]
    \includegraphics[width=.95\linewidth]{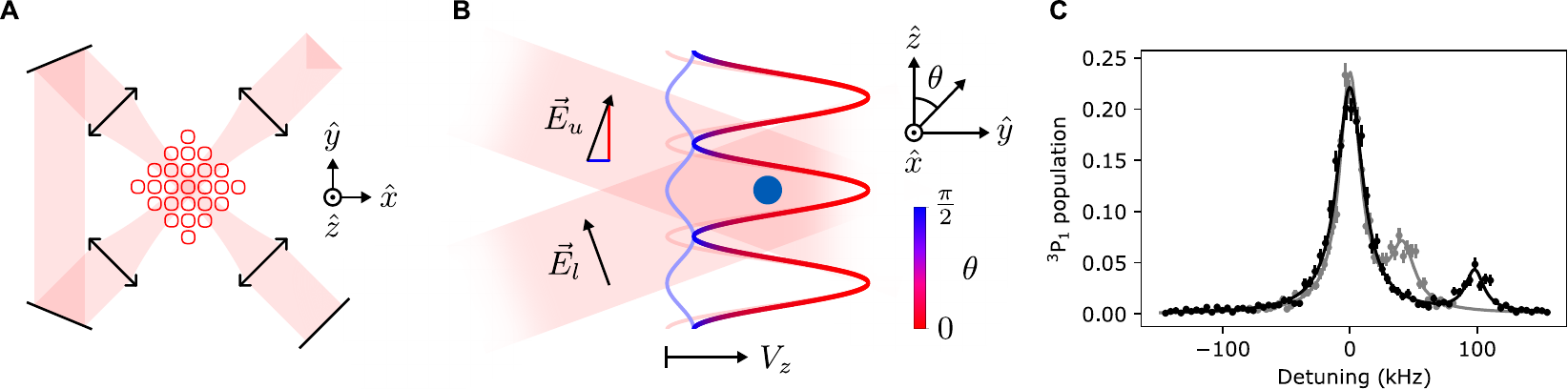}
    \caption{Resolved-sideband cooling in a magic-angle 3D optical lattice. a) Schematic of the 2D bowtie lattice, showing folding mirrors (black lines) and $4f$-relays (double arrows). Red contours denote an equipotential surface of the lattice, which is polarized along the $z$ axis. b) Schematic of the crossed-beam axial lattice, showing the polarizations $\vec{E}_u$ and $\vec{E}_l$ of the upper and lower lattice beams. Coloring denotes the polarization of the resulting standing wave, which lies in the $y$-$z$ plane making an angle $\theta$ from the $z$-axis. The two polarization components of the lattice beams result in two standing waves with orthogonal polarizations. For the shallow crossing angle of $36^\circ$ between the two beams used in this work, the depth, $V_z$, of the $z$-polarized standing wave dominates. As a result atoms (blue disk) sit at anti-nodes of this lattice, and nodes of the $y$-polarized lattice. c) In these conditions, as long as the atomic-wavepackets remain sufficiently small, the atoms experience a 3D optical lattice potential that is completely $z$-polarized, and so a robust magic-angle magnetic field condition can be found. Sideband spectra in the axial ($z$ axis, grey data) and radial ($x$-$y$ plane, black data) directions reveal that under these conditions the atoms can be cooled to average phonon occupations of $\Bar{n} = 0.000^{+33}_{-0}$ and $0.000^{+29}_{-0}$ in the two respective axes via resolved-sideband cooling, yielding a 3D ground state fraction of $100^{+0}_{-7}\%$. Each data point is averaged over 10 repetitions of the experiment using a $16\times16$ array of tweezers spanning $\rm26\times26\:\mu m$. This corresponds to a region that includes more than 2000 sites in the lattice (see Fig.~\ref{sfig:images}a).
    }
    \label{sfig:latcool}
\end{figure*}

\begin{table}[!htb]
    \centering
    \begin{tabular}{cccccc} 
     \toprule
                            & $\rm E_R$ & $\rm E_R^{Ax}$ & $\rm E_R^{2D}$ & $J_0$ & kHz ($E/h$) \\ 
     \midrule
     $\rm 1 E_R$            & - & 10.47 & 2 & 21.05 & 3.43 \\
     $\rm 1 E_R^{Ax}$       & 0.095 & - & 0.19 & 2.01 & 0.328 \\
     $\rm 1 E_R^{2D}$       & 0.5 & 5.24 & - & 10.53 & 1.72 \\
     $J_0$                  & - & - & 0.095 & - & 0.163 \\
     Axial full             & 423 & 4430 & 846 & - & 1450 \\
     Axial tunneling        & 2.48 & 26 & 4.97 & 52.3 & 8.52 \\
     2D full                & 416 & - & 832 & - & 1430 \\
     2D pre-tunneling       & 11.9 & - & 23.7 & 249 & 40.7 \\
     2D tunneling           & 2.5 & - & 5.0 & 52.6 & 8.58 \\
     Confinement start      & - & - & 1.01 & 10.6 & 1.73 \\
     Confinement end        & - & - & 0.095 & 1.0 & 0.163 \\
     Oracle                 & - & - & 1.19 & 12.55 & 2.046 \\
     \bottomrule
    \end{tabular}
    \caption{Summary of relevant energy scales. These energies $E$ are provided as a frequency $E/h$, where $h$ is Planck's constant, as well as in units of the tunneling energy $J_0$, and the recoil energies of an 813~nm photon, and the axial and 2D lattices ($\rm E_R$, $\rm E_R^{Ax}$, and $\rm E_R^{2D}$, respectively). The values provided for the depth of the oracle, and for the confinement tweezer during tunneling (confinement end) are nominal values corresponding to the data appearing in the main text. However, for specific data sets these values can differ, as explicitly mentioned in the text for relevant cases.}
    \label{tab:traps}
\end{table}

The different optical potentials used in this work can broadly be separated into two categories: tweezer arrays, which provide programmable control for state preparation and implementation of the search oracle, and the optical lattice, which provides tight confinement for optical cooling and imaging, as well as a homogeneous environment in which tunneling can occur. Most of these potentials are at the clock-magic wavelength of 813.4~nm, which decouples the $\mathrm{^1S_0} \Leftrightarrow \mathrm{^3P_0}$ optical clock transition in $\rm ^{88}Sr$ from the location of the atoms, enabling future studies that utilize this long-lived internal degree of freedom~\cite{norcia_seconds-scale_2019-1, madjarov_atomic-array_2019, young_half-minute-scale_2020} in addition to the tunneling demonstrated in this work.

One tweezer array is at a shorter wavelength of 515~nm, and is used for initial trapping and cooling of thermal $\rm ^{88}Sr$ atoms as previously described~\cite{norcia_microscopic_2018-1, young_half-minute-scale_2020}. The shorter wavelength used in this tweezer array yields a tweezer waist of 480(20)~nm, which allows these tweezers to address single sites in the optical lattice. This is used during state preparation to implant atoms into specific sites in the lattice, and during the search procedure to modify the depth of a single marked lattice site. The other tweezer array is at the clock-magic wavelength, and is as previously described~\cite{norcia_seconds-scale_2019-1, young_half-minute-scale_2020} except with a reduced aperture to increase the waist of the tweezers to 3.32~$\rm\mu m$ in order to provide a harmonic confining potential that spans many lattice sites.

The optical lattice is at the clock-magic wavelength, and is composed of a 2D bowtie lattice and a 1D crossed-beam lattice along the remaining axis (Fig.~\ref{sfig:latcool}a, b). The bowtie configuration is used in the 2D lattice for two reasons: First, since the beam is interfered four times in this configuration, it results in an eight-fold increase in lattice depth for fixed optical power in comparison to the more common configuration of two independent crossed 1D optical lattices. Second, the increased lattice constant of $a=813/\sqrt{2}$~nm in comparison to $813/2$~nm for a retro-reflected lattice is helpful for site-resolved imaging and addressing given the 440(20)~nm Gaussian waist of our imaging system, and aforementioned optical tweezer waist. This bowtie lattice is composed of a pair of $4f$-relays that are folded to cross at $90^\circ$ at their foci, where the beam waists are 55~$\rm\mu m$. To ensure that all arms of this lattice fully interfere, the polarizations of these beams are oriented out of the plane of this lattice, which we will define as the $z$ axis. With 540~mW of input power, this yields average lattice depths of $\rm 416 E_R$ (where $\rm E_R$ is the recoil energy of an 813~nm photon, see Tab.~\ref{tab:traps}), and corresponding average radial trap frequencies of 98(1)~kHz over a $\rm26\times26\:\mu m$ region. Over this region, the radial trap frequencies have a standard deviation of 5~kHz due to the finite extent of the lattice beams.

The axial lattice is composed of two parallel elliptical beams which are focused by an aspheric lens such that they cross at their foci, where the beam waists are $\rm25\times10\:\mu m$, with the 25~$\rm\mu m$ axis oriented in the plane of the 2D lattice (along the $x$ axis, see Fig.~\ref{sfig:latcool}). These beams cross with an angle of $36^\circ$, yielding a lattice constant of 1.32~$\rm\mu m$. For these beams to interfere fully their polarization should be oriented in the $x$ direction, normal to the plane in which they cross. However, this would result in a polarization that is orthogonal to that of the 2D lattice. Although it is possible to find an approximate magic-angle magnetic field~\cite{norcia_microscopic_2018-1} that is shared by both these lattice polarizations, in this situation the atoms experience very large polarization gradients. For example, when an atom moves from a node to an antinode in either the 2D or axial lattices they experience a $90^\circ$ rotation in the polarization of the light field. This results in substantial broadening of the $\mathrm{^1S_0} \Leftrightarrow \mathrm{^3P_1}$ cooling transition, and thus reduced cooling and imaging performance.

Instead, we orient the polarization of the axial lattice beams along the $z$ axis before the final focusing lens. This results in two components of the lattice with orthogonal polarizations, and spatial phases that differ by $\pi$ (Fig.~\ref{sfig:latcool}b). Because the crossing angle of our beams is well under $90^\circ$ the axial lattice is predominantly $z$-polarized. As a result, atoms sit at the anti-nodes of the $z$-polarized lattice, and at the nodes of the $y$-polarized lattice. As long as the atoms remain cold, their experience of the $y$-polarized component of the axial lattice is suppressed to first order in their position. In this regime the entire 3D optical lattice, composed of the 2D bowtie lattice and the 1D crossed-beam lattice, is effectively $z$-polarized, and a robust magic-angle magnetic field condition can be found. This enables high-fidelity resolved sideband cooling and imaging (Fig.~\ref{sfig:latcool}c), with the primary tradeoff being a slightly reduced effective lattice depth of 90\% in comparison to an $x$-polarized axial lattice. Nonetheless, with 660~mW of total optical power split between the two arms of the axial lattice we achieve average lattice depths of $\rm 423 E_R$, and average axial trap frequencies of 41(1)~kHz over a $\rm26\times26\:\mu m$ region. The axial trap frequencies have a standard deviation of 7~kHz over this region due to the more tightly focused axial lattice beams in comparison to the 2D lattice beam.

Sideband spectroscopy after resolved sideband cooling in these traps indicates average phonon occupations of $\Bar{n} = 0.000^{+33}_{-0}$ and $0.000^{+29}_{-0}$ in the axial and radial directions respectively, and thus a 3D ground state fraction of $100^{+0}_{-7}\%$  over a $\rm26\times26\:\mu m$ region corresponding to more than 2000 lattice sites. Another way of indirectly characterizing this cooling performance is via loss during tunneling, where any atom in a motionally excited state is lost either due to the lack of a bound state, or due to a significantly higher tunneling rate that causes it to leave the analysis region. The survival fraction of atoms after tunneling is typically 94\%, but can be as high as 96\% when the cooling is tuned up. When compensating for imaging loss, this suggests a 3D ground state fraction of 99\%, consistent with the theoretical maximum cooling performance given our trap frequencies and the spectrum of our cooling laser. However, this measurement is less sensitive to axial temperature and should not be taken as an accurate estimate. Moreover, because we do not tune up this cooling every day, the survival can fluctuate between 88\% and 96\% for data appearing throughout this paper. This effect can be removed by simply post-selecting data based on survival.

One challenge with the crossed-beam configuration used in the axial lattice is that the relative phase of the two beams must be carefully controlled to preserve the spatial phase of the lattice. To hand atoms from the tweezers to the optical lattice, it is critical to ensure that the anti-node of the axial lattice lines up with the foci of the optical tweezer arrays, as described in more detail in section~\ref{sec:alignment}. To ease this process we ensure that the two arms of the axial lattice are matched to within $\rm \sim100\:\mu m$ of each other, which keeps the spatial phase of the axial lattice passively stable on the several-hour timescale.

\subsection{Alignment of optical potentials}
\label{sec:alignment}

Precise alignment of the various optical potentials in this apparatus is critical for this work. As a result, it is useful to both maximize the passive stability of these potentials, and to have fast, robust signals for their alignment. To improve passive stability, we ensure that all of the optical potentials share a common reference --- namely a Macor glass-ceramic reference plate. This plate rigidly attaches to the microscope objective, which to first order defines the position of the optical tweezer arrays, and to the mirrors and relay lenses which define the position of the 2D lattice. The aspheric lens that (again to first order) defines the position of the axial lattice is also designed to attach to this reference, but due to the larger lattice spacing along this axis we have not yet found this to be necessary.

The 515~nm and 813~nm tweezer arrays are coarsely aligned to each other as previously described~\cite{young_half-minute-scale_2020}. To align the lattices to the tweezer arrays, it is useful to have a tracer beam at 688~nm that follows the same optical path as the various lattice beams. This light is sufficiently close to 813~nm to avoid significant chromatic aberrations, but pumps atoms via the $\mathrm{^3S_1}$ state from $\mathrm{^3P_1}$ to $\mathrm{^3P_{0,2}}$, which are long-lived and dark to our imaging light. As a result, these beams deplete atoms from the narrow-line magneto-optical trap (MOT) used to load atoms into the tweezers, and so the dimming of the MOT can be used to coarsely align the lattice beams. Similarly, finer alignment can be performed by maximizing loss of atoms from the tweezer array through the application of these beams. Final alignment of the lattice beams is performed by maximizing the light shift induced by the lattice beams on the $\mathrm{^1S_0} \Leftrightarrow \mathrm{^3P_1}$ transition in 515~nm tweezer-trapped atoms (in magnetic fields that are not magic for the lattice polarization). With this optimization complete, one can verify the alignment by maximizing the observed motional trap frequencies provided by the lattices, however these two procedures typically yield similar results.

Alignment of the axial lattice is particularly challenging, since the foci of two separate lattice beams need to be overlapped with the tweezer array in a $\rm25\times32\times10\:\mu m$ volume. While this can be accomplished via the above procedure, the fringes of the resulting lattice must also be flattened relative to the plane of the tweezers while maintaining this overlap. This flattening is performed in the same manner as previously reported~\cite{young_half-minute-scale_2020}, however, to ease this process we introduce an additional $4f$-relay to the axial lattice before the final aspheric focusing lens. This enables servo-actuated mirrors to be placed in both a Fourier plane and an image plane of the tweezer array. The Fourier plane mirror can be used to control a pure displacement of the lattice at the tweezer array, making it possible to compensate for drifts in beam pointing without spoiling the flattening of the lattice. This mirror is back-side polished, such that the transmission can be imaged onto a camera for in-situ monitoring of the axial lattice. The image plane mirror controls a pure tilt (rotation about $x$) of the lattice fringes, enabling flattening of the lattice in that direction. To control the roll (rotation about $y$) and spatial phase of the axial lattice, each lattice arm passes through an independently controlled optical flat. By controlling the angle of these plates, the beams can be independently displaced in the Fourier plane. Small adjustments of these plates via piezoelectric actuators can be used to change the relative phase between the lattice beams, and thus the spatial phase of the lattice. Note that in order to match the optical path length of the two lattice arms it is important that both arms include these optical flats.

As mentioned in section~\ref{sec:potentials}, the spatial phase of the axial lattice drifts on the several-hour timescale. In practice the loading, cooling, imaging, and tunneling presented in this work are relatively insensitive to this drift. As a result, this phase only needs to be adjusted when a node is close to the focal plane of the tweezers, where the transfer of atoms between the tweezers and the lattice is bistable between two layers of the axial lattice, and can result in inconsistent loading and cooling.

The spatial phase of the 2D lattice is more important, since fluctuations in the alignment between the oracle tweezer and sites in the lattice can lead to fluctuations in the effective depth of the applied oracle, or substantial overlap between the oracle and other sites in the lattice that aren't intended to be marked. The 2D lattice can also exhibit the same bistability when loading atoms as observed in the axial lattice when a preparation tweezer points in between two lattice sites. To minimize sensitivity to these drifts, especially for larger atom arrays, it is important that the lattice vectors of the tweezer array be well-matched to those of the 2D lattice. This can be done by adjusting the rotation of the tweezer array via the angle of the acousto-optic deflectors (AODs) used to project the array, and by adjusting the array position and spacing via the radio-frequency (RF) tones used to generate the array. In our apparatus one lattice spacing corresponds to a 1.95~MHz offset between 515~nm tweezers. For loading adjacent sites in the lattice it is important to keep this offset high compared to the relevant motional frequencies of the atoms, since otherwise beating between adjacent tweezers can lead to significant parametric heating and atom loss.

Even with this fine-tuning, shifts in the relative alignment between the tweezers and sites in the 2D lattice are currently the dominant source of instability in this apparatus, drifting by up to $a/2$, where $a$ is the 2D lattice spacing, over the course of an hour. This is particularly problematic when attempting to adiabatically load the lattice ground state, since loading the wrong site in the lattice corresponds to loading an excited state. This is also true of the alignment of the 813~nm confinement tweezer since, despite its larger waist, its center marks a specific site as the lattice ground state during loading, when the lattice is deep. 

We hypothesize that this instability is primarily due fluctuations in the spatial phase of the 2D lattice. Due to its bowtie design, the differential path length between the first and final passes of the beam is almost a meter (in comparison to $\rm \sim100\:\mu m$ in the case of the axial lattice). This greatly increases sensitivity to thermal fluctuations, as well as drifts in the laser frequency, where a 0.1~K change in temperature, or a 300~MHz shift in laser frequency, could result in the lattice shifting by $a/2$. Since we lock our 813~nm laser to a wavemeter with a frequency deviation sensitivity of 20~MHz, we expect the former issue to be the dominant source of instability. Luckily, these drifts are slow enough that in this work we can compensate for them via manual calibrations. In the future, using more sophisticated control software, calibration data could be automatically interleaved with experimental data to maintain this alignment.

Beyond aligning individual tweezers to sites in the lattice, for resource state preparation it is further important to align the center of the tweezer array to the center of curvature of the lattice (see section~\ref{sec:resourcePrep}). To characterize the shape of the lattice (see Fig.~\ref{sfig:tunnelfit}d), we use a light shift signal similar to that employed in the initial alignment procedure described above. However, in this case it is useful to fully extinguish the tweezers during spectroscopy, before transferring atoms back into the tweezers for readout, or performing readout directly in the lattice. This ensures that the atoms sit at the anti-nodes of the lattice, yielding consistent results, instead of being pulled away from the center of sites in the lattice due to imperfect alignment of the tweezers. For consistent ground state loading, the lattice center must be characterized in this way every few days, and the tweezer arrays shifted appropriately. 

With the lattice center identified, the preparation tweezers can be aligned by simply scanning their position, and looking for where atoms are most successfully loaded into the target sites in the lattice. This calibration must be repeated on the 30 minute-timescale due to drifts in the spatial phase of the 2D lattice. Critically, if the lattice spacing is well-characterized (see section~\ref{sec:image}), this step can be done with many atoms in a large tweezer array, offering higher statistics and fairly fast ($\lesssim 1$ minute) calibration. Contrarily, the alignment of the confinement tweezer is fairly time-consuming, since only a single tweezer is involved. In this case, we attempt to adiabatically load the ground state of the shallow lattice, as described in the main text, hold for 10~ms~$=10.26\tau$, and reverse the ramp as a function of tweezer position. If the tweezer is well-aligned, this results in 57(5)\% of the atoms returning to where they were initially implanted in the lattice. If poorly aligned, this ramp is no longer adiabatic and does not load an eigenstate of the lattice. In this case, the atomic wavefunction evolves during the hold such that the reverse ramp leaves the atom at a different location (or simply results in loss of the atom). Luckily, this time-consuming protocol only has to be performed once every few days when the position of the lattice center moves significantly. This is because the relative alignment of the two tweezer arrays is extremely stable, at least as monitored via a pickoff directly before the objective, remaining within $a/20$ of each other over multiple days. As such, the results of the preparation tweezer alignment can be fed-forward to the confinement tweezer to maintain its alignment. As long as the lattice center remains in the same position, this procedure yields equivalent performance to the full alignment procedure. While there could be fluctuations in the tweezer positions downstream of the pickoff used to monitor them, if such fluctuations exist we expect them to primarily be common mode and not differential, both on physical grounds, and based on the robustness of this feed-forward procedure. However, common-mode fluctuations in the tweezer positions could also contribute to the observed 2D lattice-tweezer alignment instability.

\subsection{Experimental sequence}
\label{sec:sequence}
\begin{figure*}[!t]
    \includegraphics[width=\linewidth]{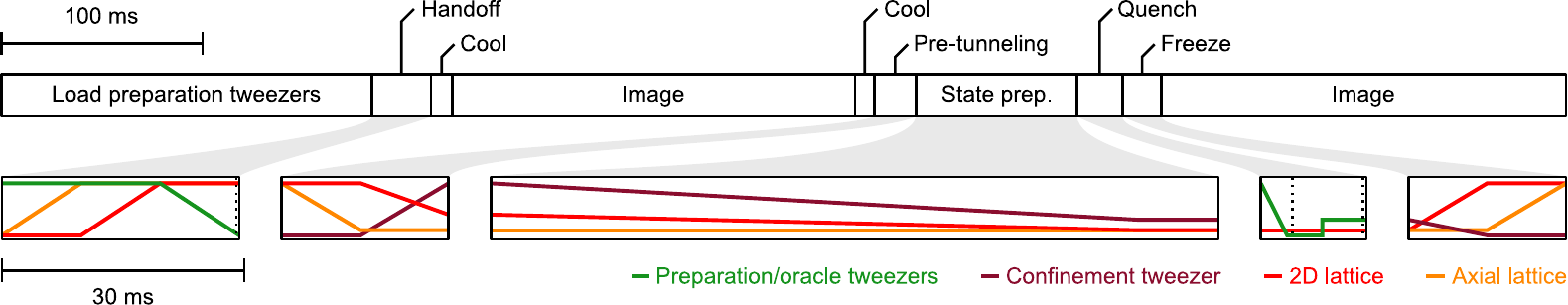}
    \caption{Timing diagram of a typical experimental sequence. The cycle time of these experiments is dominated by the initial loading of thermal atoms into the preparation tweezers, as well as the pair of relatively long images used to detect the atom positions (see text). Callouts show the intensity of the various optical potentials during different phases of the experiment, including, from left to right, handing atoms from the preparation tweezers to the 3D lattice, ramping to shallow lattice conditions in preparation for tunneling, adiabatically preparing the ground state of the shallow lattice, quenching to the search Hamiltonian, and freezing the atoms in place after their evolution in the lattice. The vertical scale in these callouts is arbitrary, but the bottom of the scale corresponds to fully extinguished beams. The dotted vertical lines denote when the shutter in the preparation tweezer system opens or closes (see text).
    }
    \label{sfig:sequence}
\end{figure*}

The timing diagram for a typical experimental sequence is shown in Fig.~\ref{sfig:sequence}. To prepare individual atoms in the optical lattice we load thermal $\rm ^{88}Sr$ atoms into the preparation tweezers and optically cool those atoms to near their 3D motional ground state as previously described~\cite{norcia_microscopic_2018-1, young_half-minute-scale_2020}. This cooling does not have to be particularly high fidelity, since the aim is simply to shrink the atomic wavepacket to be substantially smaller than a single site in the lattice. At this point we perform a sequence of ramps, each 10~ms long, to transfer the tweezer-trapped atoms into the lattice. The axial lattice is ramped up first, followed by the 2D lattice, and finally the preparation tweezers are ramped down. Once the tweezers are ramped low we fully shutter the tweezer beam, and then ramp the beam back up to full power. This allows us to avoid thermal lensing effects that can cause pointing fluctuations in the tweezers, while also avoiding unwanted light shifts introduced by the tweezers. Once the atoms are in the optical lattice, we switch from a magic magnetic field condition for the 515~nm tweezers~\cite{norcia_microscopic_2018-1} to a magic field condition for the 813~nm optical lattices~\cite{norcia_seconds-scale_2019-1} and perform resolved sideband cooling. This cooling is composed of 50 pairs of 100~$\rm \mu s$-long pulses of cooling light alternating between the axial and transverse directions. This brings $100^{+0}_{-7}\%$ of the atoms to their 3D motional ground state, as diagnosed through sideband spectroscopy (Fig.~\ref{sfig:latcool}c). For experiments like those in this work that involve only single atoms, this cooling performance is not critical and manifests only as less efficient data collection. In future studies involving multiple particles that must all be cooled to indistinguishability, this cooling could be further improved by increasing the lattice depth, improving the stability of the various optical potentials, and via more advanced cooling schemes that are not limited by the intrinsic linewidth of the $\mathrm{^1S_0} \Leftrightarrow \mathrm{^3P_1}$ transition in strontium~\cite{kaufman_cooling_2012, brown_hyperpolarizability_2017-1}.

To detect atoms in the lattice, we simply turn on the two cooling beams in addition to a beam that off-resonantly scatters off of the $\mathrm{^1S_0} \Leftrightarrow \mathrm{^1P_1}$ transition. This scattered light is collected and imaged onto an EMCCD camera as previously described~\cite{norcia_microscopic_2018-1, young_half-minute-scale_2020}. Note that in the lattice it is particularly important to keep the atoms near their motional ground states throughout the duration of the imaging step, since the tunneling rate can be very rapid in higher bands, leading to perceived atom loss. As a result, at typical lattice depths the duration of each image is 200~ms to ensure that the cooling rate outpaces the heating rate due to scattering, yielding an combined imaging infidelity and loss of 3.5(3)\%. The imaging performance is expected to improve dramatically with increased lattice depth since increasing the number of bands that can be occupied without appreciable tunneling would enable faster cooling schemes and thus shorter, higher fidelity images~\cite{covey_2000-times_2019, norcia_seconds-scale_2019-1}. To achieve this, we have been exploring alternative laser sources at 813~nm~\cite{eckner_high-power_2021}.

After identifying atoms in the lattice via an initial round of imaging, we cool the atoms once more as previously described, and ramp the axial and 2D lattices to $\rm26.0 E_R^{Ax}$ and $\rm23.7 E_R^{2D}$ respectively (see Tab.~\ref{tab:traps} for units), again with sequential, 10~ms long ramps. Where relevant, the confinement tweezer is ramped up as the 2D lattice ramps low. At these depths the tunneling rate is still negligible, but this serves as a good starting point for adiabatic ramps of the tunneling energy. Note that tunneling in the axial direction is frozen out in all regimes we operate in by the detuning between different sites in the axial lattice of $\Delta = 2.84$~kHz due to gravity.

Subsequent operations are dependent on the specific experiment we are interested in performing. For the 2D continuous-time quantum walks, we simply quench the depth of the 2D lattice to $\rm5.0 E_R^{2D}$, where the average tunneling rate is $J_0/\hbar = 2\pi\times 163$~Hz. To prepare the lattice ground state, we perform the ramp described in the main text, composed of an 80~ms-long linear ramp of the 2D lattice depth from $\rm23.7 E_R^{2D}$ to $\rm5.0 E_R^{2D}$, and of the confinement tweezer depth from $10.6J_0$ to $1.0J_0$. For the spatial search data shown in the main text, to quench to the oracle Hamiltonian we servo the tweezer intensities low and reprogram their positions to implement the appropriate oracle. At this point the shutter is opened, and the tweezer intensity jumped to the appropriate value. 

The tweezer intensity servo has a relatively slow settling time of 0.7~ms. Based on our calculations, this is not expected to fundamentally change the behavior of the experiment. To confirm this, we have also performed the quench by using a sample and hold feature on this intensity servo. In this case, while the shutter is closed, the tweezers are moved and servoed to the appropriate locations and depths. The sample and hold feature is then engaged, and the tweezers fully extinguished via an RF switch. The shutter is then opened, and the tweezer switched on and off during the quench with a much faster rise and fall time of $150$~ns. While this sample and hold procedure yields slightly less stable oracle depths than in the above case, under these conditions we still observe similar search dynamics as shown in the main text. The data appearing in Fig.~\ref{sfig:images}fg is generated using this modified quench procedure.

Because the desired oracle depths are approximately four orders of magnitude shallower than the depth of the preparation tweezers (see section~\ref{sec:simulations}), it is challenging to design a feedback loop on the tweezer intensity that has sufficient dynamic range. Instead, we make 24 extra tweezers away from the atoms where a total of 9 tweezers are at the oracle depth, 12 are $\sim24\times$ deeper than the oracle, and 4 are $\sim570\times$ deeper than the oracle. This makes it easy to servo the central tweezer to the appropriate depth by controlling the total power in the array~\cite{brown_gray-molasses_2019}. 

At the end of these experiments, we freeze the atoms in place by jumping the 2D lattice depth back to $\rm23.7 E_R^{2D}$. The two lattices are then ramped back up to full depth in reverse order of when they were ramped shallow, after which an additional image reads out the final positions of the atoms. The full cycle time of these experiments is approximately 700~ms, and is dominated by the two 200~ms-long images, as well as the 150~ms-long loading sequence. As discussed above, the imaging time can potentially be reduced substantially with increased lattice depth. The loading sequence could also be bypassed by recycling atoms~\cite{norcia_seconds-scale_2019-1} using the preparation tweezers, either to return atoms to the appropriate locations after each run of the experiment, or to move atoms from a large reservoir of tweezer-trapped atoms to the experiment region. Such improvements could result in substantially faster cycle times approaching 100~ms.

\subsection{Image analysis}
\label{sec:image}

\begin{figure*}[!t]
    \includegraphics[width=.9\linewidth]{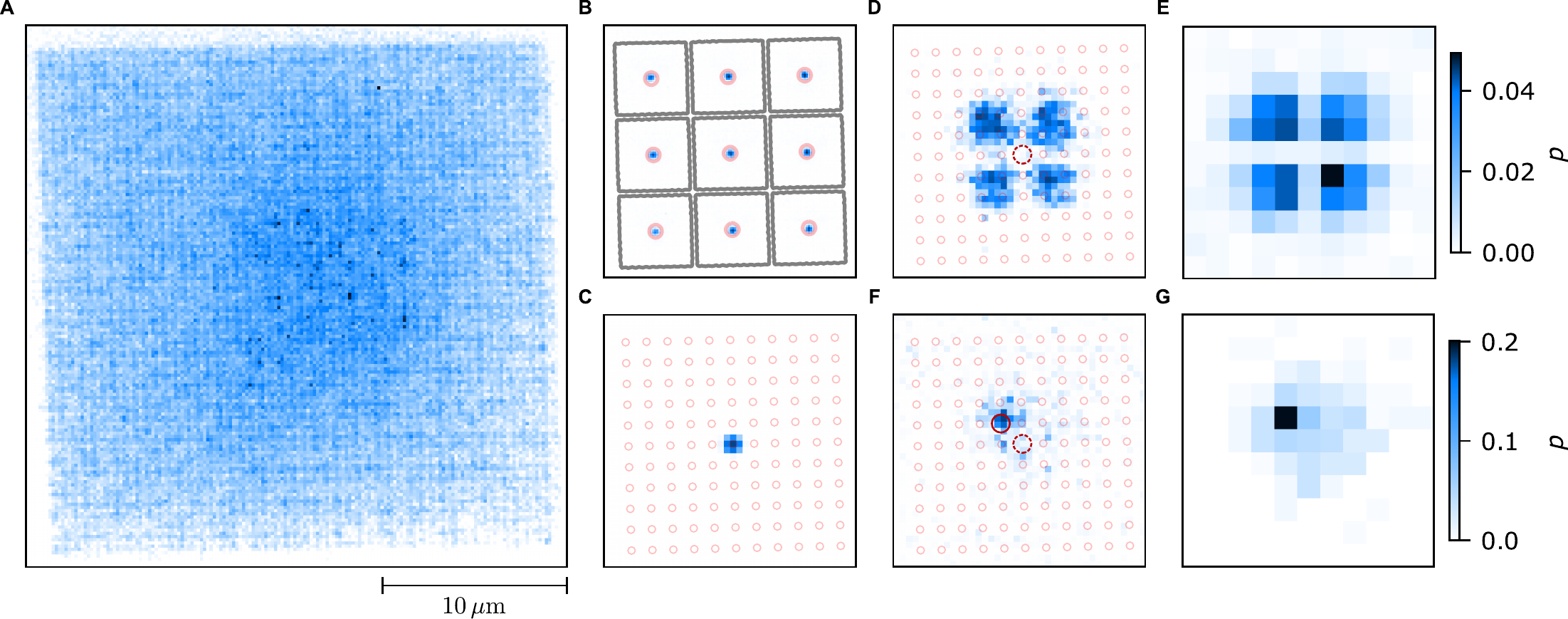}
    \caption{Image analysis. a) An averaged image of atoms loaded into $\sim2500$ sites in the lattice is used to calibrate the positions of these sites for subsequent analysis. Note that only a small subset of these sites are loaded on any individual run of the experiment. b) For data to the left of the red line in Fig.~\ref{fig:setup}c in the main text, we implant atoms in nine sites in the lattice, shown here as an averaged image. We define regions of the lattice (black contours) surrounding each preparation tweezer, where for specific evolution times the atoms will remain in the region they started in. To identify implanted atoms, we integrate the product of single-shot images with a mask defined by the known point spread function (PSF) of the imaging system at the corresponding locations in the lattice (red contours denote where these masks fall to 10\% of their maxima). c) To study quantum walks of individual atoms, we overlay the different regions, averaging over both repetitions of the experiment and regions in the lattice. Red circles denote the independently calibrated locations of sites in the lattice. d) After allowing the atoms to propagate (in this case for 2.75~ms~$=1.3\tau$) the atomic wavefunction coherently spreads across the lattice. Images averaged over repetitions and over the nine regions in this example constitute a measurement of the atomic probability density. e) To fully characterize this probability density $p$ we identify which lattice site the imaged atoms occupy (see text), and generate a histogram of those locations. f) Averaged images of exemplary spatial search data (using only a single region) show the atoms move from the initial site the atom was implanted in (dark red dashed circle) to the site that the oracle was applied to (dark red circle) after an evolution time of 10~ms following the quench. g) Corresponding probability densities for exemplary spatial search data. Note that the oracle depth used in this data is $ 3.1 J_0 $, yielding slightly higher transfer than for the deeper oracle used in the main text. The color scales in (a, b, c, d, f) are arbitrary, and for visualization purposes only. Each data set is averaged over $\sim 300$ repetitions.
    }
    \label{sfig:images}
\end{figure*}

To identify the atom locations while minimizing the required signal and thus imaging time, we take advantage of a few extra pieces of information. First, the atoms can only occupy a discrete set of sites in the lattice, and the locations of these sites can be characterized through independent measurements (Fig.~\ref{sfig:images}a). For each site in the lattice, we multiply the raw images by a mask corresponding to the measured point spread function (PSF) of the imaging system at that location (Fig.~\ref{sfig:images}b-g) and sum the result to obtain our signal. This procedure can be thought of as a version of image deconvolution which takes advantage of the additional fact that bright points in the underlying image should come from only a discrete set of locations~\cite{sherson_single-atom-resolved_2010}.

Additionally, in the first image, atoms are loaded into a known subset of sites in the lattice corresponding to where the preparation tweezers were. We can keep these loaded sites far enough apart such that no atoms cross from the region surrounding one preparation tweezer to another (Fig.~\ref{sfig:images}b). When an atom is identified in the first image, we analyze only the brighest signal in the corresponding region in the second image, removing intermediate signals due to the overlapping PSFs from different sites in the lattice. Based on Monte-Carlo simulations, this procedure results in an error rate for misidentifying the location of an atom that is bounded to below 2.5\%.

\subsection{Characterizing tunneling}

\begin{figure*}[!t]
    \includegraphics[width=\linewidth]{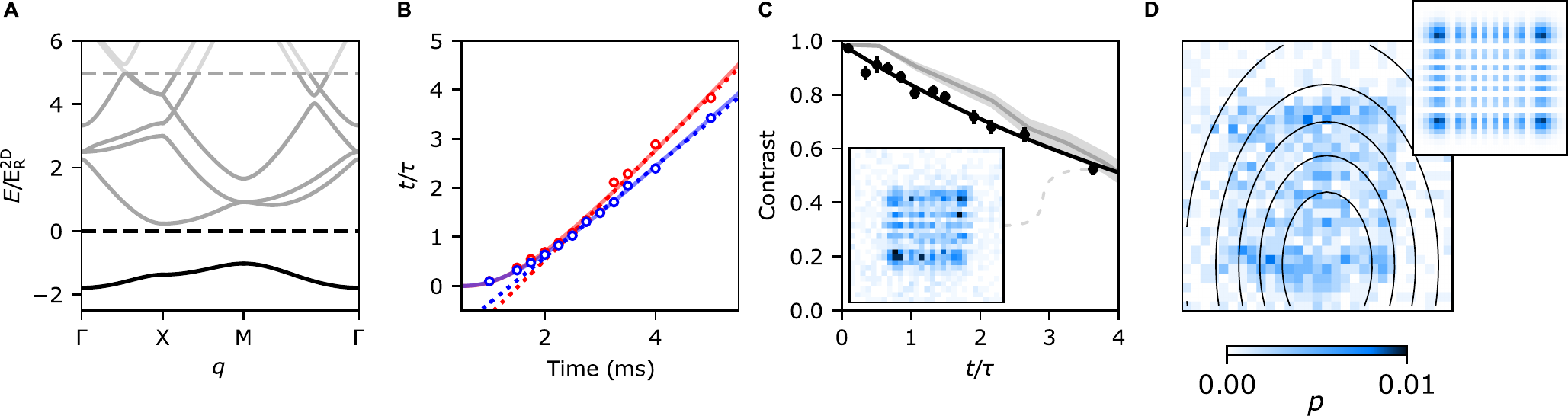}
    \caption{Characterizing 2D continuous-time quantum walks. a) Band structure plotted along a triangular path between critical points at the center, edge center, and corner of the first Brillouin zone ($\rm \mathsf \Gamma$, $\rm \mathsf X$, and $\rm \mathsf M$ respectively), where $q$ is the quasimomentum and $E$ the energy. $k$ and $\rm E_R^{2D}$ are the wavevector and recoil energy of the 2D lattice, respectively, and the energy scale is set so that bound states in the 2D lattice have negative energy (below dashed black line). For the lattice depths used in this work, there is only a single bound band (black line) in the 2D lattice. However, there are multiple bands (dark grey lines) that are not bound by the 2D lattice, but are lower in energy than the axial lattice depth (dashed grey line). This can result in imperfect filtering of atoms that are parametrically heated to higher bands, but do not leave the analysis region. b) The atomic probability density distributions are measured as a function of evolution time (see Fig.~\ref{fig:setup} of main text). At each evolution time, these distributions are fit to a model that allows for independent values of the unitless tunneling times $t_x/\tau$ and $t_y/\tau$ in the $x$ and $y$ directions (red and blue circles respectively). At late evolution times, these values grow linearly with time corresponding to constant tunneling energies $J_x/\hbar = 2\pi\times 178.1(1.1)$~Hz and $J_y/\hbar = 2\pi\times 148.3(1.1)$~Hz in the two axes (dashed lines are linear fits excluding the first 4 points). At early times these values deviate from this linear expectation due to the slow settling time of the lattice intensity (see text), which is captured by hyperbolic fits (solid lines) which agree with the linear expectations at late times.  c) The contrast of these fitted fringes decays with an exponential $1/e$ time constant of $6.1(3)\tau$. This decay is in reasonable agreement with a Monte-Carlo simulation that includes the measured lattice curvature, and random stray reflections at the $6\times10^{-6}$ level in intensity in comparison to the lattice beams (grey line, shaded region denotes $1\sigma$ confidence intervals), however, other effects could also contribute to this decay (see text). Callout shows the probability density corresponding to the final point (same data as in Fig.~\ref{fig:setup}b of the main text), which still exhibits clear interference fringes. d) At later times, in this case at $5.2\tau$, the atoms begin to sample the curvature of the lattice, and the probability density deviates substantially from the symmetric prediction in a flat lattice (inset). This behavior is qualitatively consistent with the expectation given the independently characterized lattice potential (contours denote equipotential surfaces of the lattice). Color scale is shared between (c) and (d), and the data appearing in (d) is averaged over 6000 repetitions of the experiment.
    }
    \label{sfig:tunnelfit}
\end{figure*}

For the tunneling conditions in this work, we operate at a fairly shallow 2D lattice depth of $\rm5.0 E_R^{2D}$. Due to this shallow depth, and the bowtie design of our lattice, it is necessary to solve for the band structure of the system numerically (Fig.~\ref{sfig:tunnelfit}a). Based on these calculations, we conclude that the effect of diagonal and next-nearest-neighbor tunneling is negligible for all results presented in this work. At these shallow depths, there is only a single bound band in the 2D lattice. However, the presence of the axial lattice can lead to trapping of states that aren't directly trapped in the 2D lattice, and closely resemble free particle states. While these bands are not typically occupied directly after loading atoms into the lattice, the procedure for adiabatic resource state preparation can result in atoms occupying these higher bands, contributing to a background signal. We find that intentionally misaligning the lattice center from the confinement tweezer can result in a filtering effect, where ground-band atoms are loaded into the local ground state defined by the confinement tweezer, whereas atoms occupying higher bands can leave the the confinement tweezer and sample a much larger region of the lattice that is outside of our analysis region, and thus interpreted as loss.

To directly characterize the tunneling rate in the lattice, we fit the measured probability density distributions as a function of tunneling time to propagation under the Hamiltonian defined in Eqn.~\ref{eq:ham} of the main text, assuming a perfectly flat lattice with constant $V_i$. The free parameters in this fit include independent values of the unitless tunneling times $t_x/\tau$ and $t_y/\tau$ in the $x$ and $y$ axes, as well as an overall contrast and offset (Fig.~\ref{sfig:tunnelfit}bc). The offset does not evolve appreciably over the evolution times explored in this work, and is consistent with overall error rates due to imperfect imaging and cooling. The data exhibits constant values of the tunneling energy at late times, reflecting the expected ballistic expansion of the atomic wavefunction. At early times, the data deviates from this linear expansion due to the relatively slow 0.7~ms settling time of the intensity servo used to control the 2D lattice depth, which results in artificially low tunneling rates. By either fitting this evolution with a hyperbola, or fitting a line to only late-time data, we find that the lattice has tunneling energies of $J_x/\hbar = 2\pi\times 178.1(1.1)$~Hz and $J_y/\hbar = 2\pi\times 148.3(1.1)$~Hz in the $x$ and $y$ directions respectively. This is in good agreement with the calculations described above, and independent characterizations of the lattice beams and lattice vectors (Fig.~\ref{sfig:images}a). Note that the discrepancy between the two axes is because, as a result of our alignment procedure, the lattice is not perfectly square.

We do not observe any atom loss over the evolution times explored in this work. However, the fitted contrast of these quantum walks decay with an exponential $1/e$ time constant of $6.1(3)\tau$. We hypothesize that this is due to inhomogeneity in the lattice, both due to laser speckle and lattice curvature, which causes the walks to deviate from the expectation in a perfectly flat lattice. A Monte-Carlo simulation which includes the independently characterized lattice curvature, and Gaussian-distributed laser speckle with a standard deviation of $6\times10^{-6}$ in comparison to the lattice beam intensity, matches the observed decay reasonably well at late times (Fig.~\ref{sfig:tunnelfit}c). We suspect that the discrepancy at early times is related to the 0.7~ms settling time of the lattice intensity servo, which could result in imperfect conversion between real units of time and the effective tunneling time. However, this discrepancy could also be the result of additional, as of yet unidentified decoherence mechanisms. At late evolution times of $5.2\tau$ (Fig.~\ref{sfig:tunnelfit}d) the walk differs significantly from the expectation in a flat lattice, but exhibits a few interesting features. First, the atomic probability density still exhibits clear interference fringes, suggesting that any disorder in the lattice is static as a function of time. Secondly, we see that the bottom edge of the atomic wavefunction has continued to expand, whereas the top and sides are beginning to oscillate back towards the center. This is qualitatively consistent with the independently characterized position and shape of the lattice potential, suggesting that the evolution of these quantum walks remains unitary even at these late times.

\subsection{Resource state preparation}
\label{sec:resourcePrep}

\begin{figure}[!t]
    \includegraphics[width=\linewidth]{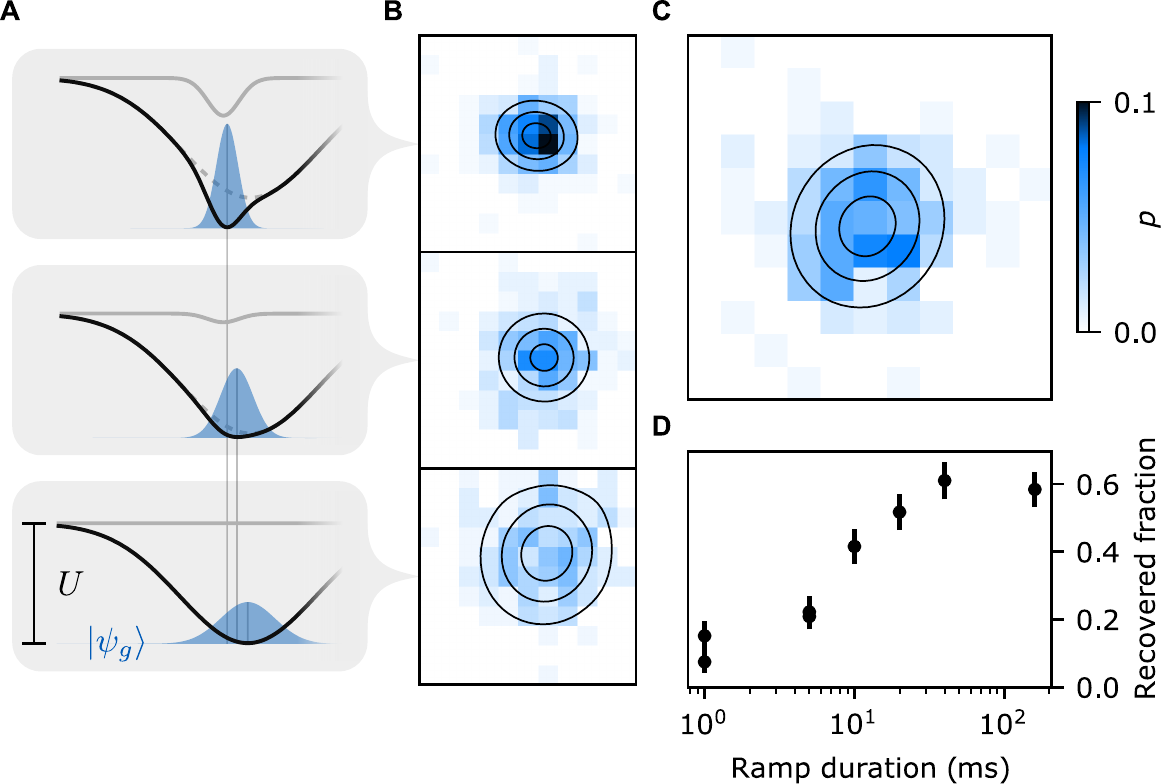}
    \caption{Self-centering ground state preparation. a) Cartoons of how the combination of misalignments between the confinement tweezer (grey curves) and lattice center (dashed grey curves), and changes in their relative depth, can lead to a shift in the minimum (vertical grey lines) of the overall potential experienced by the atoms (black curves), and a corresponding change in the size and position of the atomic ground state (blue shaded regions). These diagrams are for visualization purposes only, and are not to scale. Furthermore, they should be understood as an average envelope spanning many lattice sites, where modulations on the scale of the lattice period have been omitted for clarity. b) With poor centering between the confinement tweezer and lattice center, measurements of the atomic probability density after performing an adiabatic ramp to the ground state reveal states whose position (and size) vary with the final confinement tweezer depth. From top to bottom, these depths are: $2.2J_0$, the typical depth used for resource state preparation of $1.0J_0$, and $0.0J_0$. c) Typical atomic probability density in the resource state, with the confinement tweezer well-aligned to the lattice center. This data corresponds to the conditions used in the main text, and appears in Fig.~\ref{fig:quench}c. Color scale is shared between (b) and (c), and each data set in these sub-figures is averaged over 600 repetitions. d) As an additional test of the resource state preparation, we perform a forward ramp to the resource state, and then a reverse ramp with variable duration. For longer ramps we recover 57(5)\% of the atoms, whereas for shorter ramps the evolution is diabatic, and very little of the atomic probability density returns to the central site, beyond what was already present on that site in the resource state. Each point in this sub-figure is averaged over 200 repetitions.
    }
    \label{sfig:ground}
\end{figure}

The procedure used to prepare the resource state in the main text is particularly sensitive to the alignment of the confinement tweezer to the center of the lattice. This is because during the adiabatic ramp the relative depths of these potentials change. If their centers are in different locations, this also results in a shift in the minimum of the combined potential experienced by the atoms. For large misalignments, this causes the state preparation to fail. However, for smaller misalignments of $\lesssim5$ lattice spacings, the center position shifts adiabatically during the ramp, leading to the successful loading of a ground state that is displaced from the initial site that the atom was loaded in (Fig.~\ref{sfig:ground}ab). This effect must be carefully controlled during the search procedure when defining the lattice origin, and exploring search as a function of oracle position. In this work, we do this by centering the preparation and confinement tweezers on the combined center of the lattice as described in section~\ref{sec:alignment}, and characterizing the prepared resource state to ensure that it is centered (see Fig.~\ref{sfig:ground}c) before proceeding with spatial search.

Despite being an inconvenience in this work, the robustness of this loading procedure against misalignment could be a useful tool. Because the alignment between the tweezers is exceptionally stable (see section~\ref{sec:alignment}), the confinement tweezer can be used as an intermediary between the preparation tweezer and the lattice, allowing one to load the lattice ground state even when the tweezers are significantly misaligned from the lattice center (Fig.~\ref{sfig:ground}b). Specifically, an appropriate ramp in the depths of the confinement tweezer and 2D lattice results in a self-calibrating ramp of the center of the combined optical potential formed by the tweezers and the lattice (Fig.~\ref{sfig:ground}a). This works even when the confinement tweezer is fully extinguished at the end of the ramp, in which case the center position starts at the location of the implanted atom, and ends at the location of the lattice center. The resultant state (Fig.~\ref{sfig:ground}b, bottom) behaves similarly to the resource state presented in the main text: it does not significantly evolve over time, and 50(14)\% of the atom population can be recovered in the initial site when reversing the ramp. Critically, we measure no atoms in the initial site after a similar set of ramps without the confinement tweezer.

Since the axial lattice depth is held constant during these ramps, a similar shift in the combined center of the axial and 2D lattices can occur. We have attempted to mitigate this by ramping the axial and 2D lattices together, maintaining a constant ratio, but this places more stringent requirements on the adiabaticity of these ramps, and results in poorer overall performance. Instead, because these potentials have significantly less curvature than the confinement tweezer, we find it sufficient to align the tweezers to the combined center of the lattices at the final ratio used during tunneling and spatial search.

While we do not do full state tomography to characterize the prepared ground (or resource) state, there are several additional pieces of evidence indicating the successful preparation of this state beyond what is described in the main text. First, the size and position of the prepared state is consistent with expectations based on independent characterizations of the various optical potentials in the experiment. For example, the data in Fig.~\ref{fig:ground}b upper and middle are taken back to back, with final confinement tweezer depths of $2.2J_0$ and $1.0J_0$ respectively. In this case we expect a ratio in the $1/e^2$ radius of the prepared states of 0.78, and observe 0.83(11). Secondly, we expect forwards and reverse ramps to recover atoms in the central site only when the evolution is adiabatic, and genuinely prepares the lattice ground state. When performing these ramps, the shorter the duration of the reverse ramp, the fewer atoms return to the central site (Fig.~\ref{fig:ground}d), as expected. The same is also true when both the forward and reverse ramp times are varied. Moreover, when the prepared state is perturbed by a quench with a centered oracle, we expect, and observe, that the reverse ramp also fails despite the fact that immediately after the quench there is more population on the central site than is present after only the forward ramp. This is because the quench results in evolution to a state that is not adiabatically connected to the ground state of the lattice when tunneling is turned off, where the atom sits on the central site.

\begin{figure}[!t]
    \includegraphics[width=.75\linewidth]{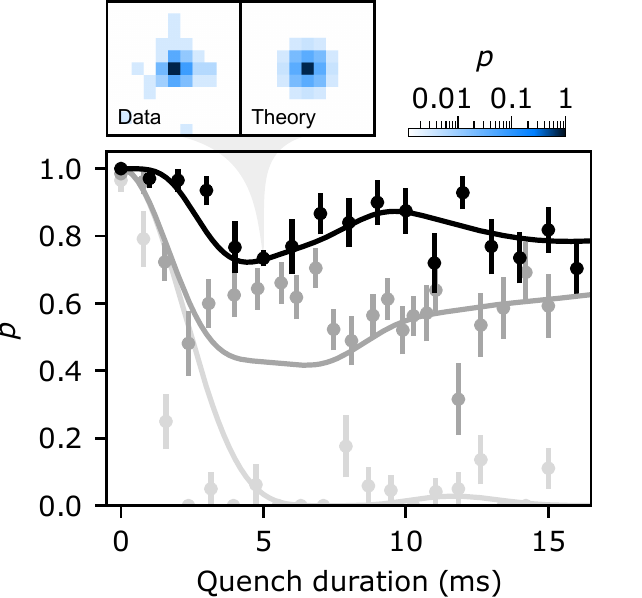}
    \caption{Resource state preparation by reversing the search algorithm. By implanting an atom in the central lattice site, and quenching with an oracle at that site with a depth of $6.2J_0$, we observe coherent oscillations in the atomic probability density (black points) on this site due to the same dynamics as in the search algorithm. Callout: characterizing the atomic probability density as a function of position at the first minimum in this oscillation (data, left), we observe the expected behavior where much of the atom population remains on the central site, but some has oscillated into a superposition that is qualitatively similar to the desired resource state (theory with no free parameters, right). For comparison, similar quenches are performed with an oracle depth of $4.1J_0$ (dark grey points), and without the oracle (light grey points), showing the atom simply leave the central site, and not return on the timescales explored in this work. Theory curves with no free parameters match the deeper quench well, and are in reasonable agreement with the shallower quench, whose behavior is more sensitive to fluctuations in the depth of the applied oracle. Each point in this figure is averaged over 60 repetitions, except the point corresponding to the callout, which is averaged over 600 repetitions.
    }
    \label{sfig:reverse}
\end{figure}

As mentioned in main text, an alternative route to preparing this resource state is to run the search procedure backwards, ``hiding'' the atom in the lattice ground state (Fig.~\ref{sfig:reverse}). By starting with the atom localized to a single site, and quenching with an oracle at that position, the atom should oscillate into the resource state. To improve the fidelity of this procedure, we operate at a somewhat lower oracle depth of $6.2J_0$, leading to slower and slightly larger amplitude oscillations than shown in the main text. When reducing the oracle depth even further, to $4.1J_0$, this procedure yields less clearly sinusoidal behavior, and is more sensitive to fluctuations in the applied oracle depth. The behavior of these quenches is in good agreement with theory with no free parameters, although the increased sensitivity to fluctuations at the intermediate oracle depth results in less accurate agreement with theory. Note that in these experiments, since the adiabatic state preparation step is omitted, the theory curves contain only information about the independently characterized optical potentials. However, even when operating at these shallower oracle depths, due to the limitations of performing this procedure in a 2D square lattice, the resultant fidelity is significantly worse than the adiabatic protocol. Nonetheless, this measurement serves as an additional confirmation of the dynamics contributing to our demonstration of spatial search, and, by extension, that the adiabatic protocol is indeed preparing the appropriate resource state. These dynamics also suggest that in a more fully connected lattice, this protocol could be a viable way of preparing the resource state with identical scaling to the search procedure, removing any additional overhead associated with the adiabatic preparation of this state. 

\section{Level structure and boundary conditions in the search Hamiltonian}
\label{sec:simulations}

\begin{figure*}[!t]
    \includegraphics[width=.95\linewidth]{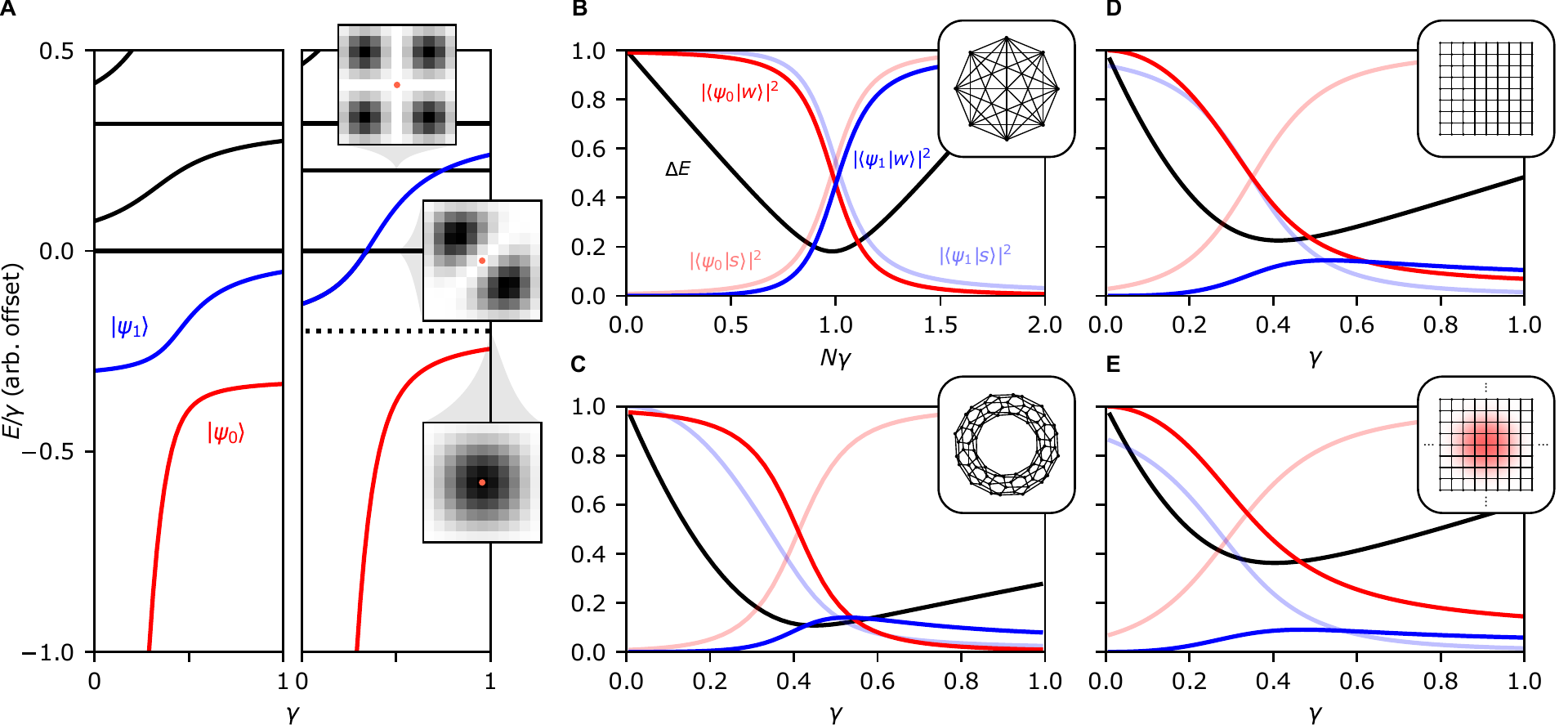}
    \caption{Level structure of the search Hamiltonian for various graphs. a) Exploring the level diagram of the search Hamiltonian as a function of the relative strength $\gamma$ of the diffusion and oracle terms, we identify a subspace spanned by states $\ket{\psi_0}$ and $\ket{\psi_1}$. This subspace is primarily supported by the ground state of the diffusion Hamiltonian on the graph, $\ket{s}$, and the marked state, $\ket{w}$ (see text). In the case of a lattice with cyclic boundary conditions (left), this subspace corresponds to the ground (red) and first excited (blue) states. In the case of closed boundary conditions (right) and for a Gaussian confining potential (not depicted), the eigenstates of the lattice have spatial structure, resulting in states (top and middle callouts) that do not overlap with a given oracle (positions marked by the red dots). In this case $\ket{\psi_0}$ and $\ket{\psi_1}$ (red and blue) need not be the ground and first excited states, and are instead identified by their overlap with $\ket{s}$ (dotted line and lower callout) and $\ket{w}$. The behavior of $\ket{\psi_0}$ and $\ket{\psi_1}$ vary for different graph geometries, both in terms of their overlap with $\ket{s}$ and $\ket{w}$, and their energy separation $\Delta E$. Insets in (b-e) denote the relevant graph connectivity and boundary conditions, but are for visualization purposes only and are not accurate in terms of graph size. These conditions are: b) A fully connected graph with $N=121$ nodes, as applies to the case of continuous-time Grover's search. c) A 2D square lattice with cyclic boundary conditions and $N=11\times11$. d) A 2D square lattice with closed boundary conditions and $N=11\times11$. e) A 2D square lattice with an additional Gaussian confining potential with a waist of 5.8 lattice sites and a depth of $1.0J_0$, as provided by the confinement tweezer during the search procedure in the main text. For (b, c), translational symmetry dictates that the behavior of the system is independent of oracle position, whereas in (d, e), the behavior is dependent on oracle position, and we plot the special case of a centered oracle.
    }
    \label{sfig:bands}
\end{figure*}

To understand the asymptotic scaling of continuous-time quantum walk-based search algorithms, it is instructive to solve for the spectrum of the search Hamiltonian as a function of the relative strength of the diffusion and oracle terms. In this section, we adopt the notation of Childs and Goldstone~\cite{childs_spatial_2004-1}, and rewrite the diffusion and oracle Hamiltonians as:

\begin{align}
    H_{lat} &=-\gamma A \nonumber \\
    H_w &=-\ket{w}\bra{w}
\end{align}

\noindent respectively, where $A$ is the adjacency matrix of the graph that the quantum walk occurs on, and $\ket{w}$ is the marked site. The parameter $\gamma$ takes on a similar role to $V_w$ in the main text, setting a relative energy scale between the diffusion and oracle terms in the full search Hamiltonian $H = H_{lat}+H_w$.

In Fig.~\ref{sfig:bands} we study the spectrum of $ H $ as a function of $\gamma$ for several different graphs. In each case we define the resource state, or equivalently the ground state of $H_{lat}$, to be $\ket{s}$. In the typically studied case of cyclic boundary conditions, the eigenstates of the lattice are plane waves, and so all eigenstates have equal overlap with the oracle, and are affected by varying $\gamma$ (Fig.~\ref{sfig:bands}a). In this setting, the search algorithm can be interpreted as the formation of a subspace composed of the ground and first excited states of the system, $\ket{\psi_0}$ and $\ket{\psi_1}$, which is mostly supported by the states $\ket{s}$ and $\ket{w}$.

This analysis helps to reveal the desirable properties associated with continuous-time Grover's search~\cite{farhi_analog_1998}, which is equivalent to search by continuous-time quantum walk on a fully connected graph (Fig.~\ref{sfig:bands}b). At the critical value of $\gamma= 1/N$ (for large $N$), where $N$ is the size of the search space, $\ket{\psi_0}$ and $\ket{\psi_1}$ have approximately equal contributions from $\ket{s}$ and $\ket{w}$. This overlap sets the contrast of the oscillations between $\ket{s}$ and $\ket{w}$ during the search procedure, whereas the energy gap $\Delta E$ between these states sets the frequency of these oscillations. For a 2D square lattice with cyclic boundary conditions (Fig.~\ref{sfig:bands}c) the ground state $\ket{\psi_0}$ undergoes a similar crossover from $\ket{w}$ to $\ket{s}$ at a critical value of $\gamma = \frac{1}{4\pi}\ln{N}+0.0488$~\cite{childs_spatial_2004-1}, however, at this critical value the overlap between $\ket{\psi_1}$ and $\ket{w}$, as well as the value of $\Delta E$, are substantially reduced relative to the case of a fully connected graph. This results in lower-contrast and lower-frequency oscillations during the search procedure, and thus less favorable asymptotic scaling of the search algorithm with $N$.

\begin{figure}[!t]
    \includegraphics[width=.95\linewidth]{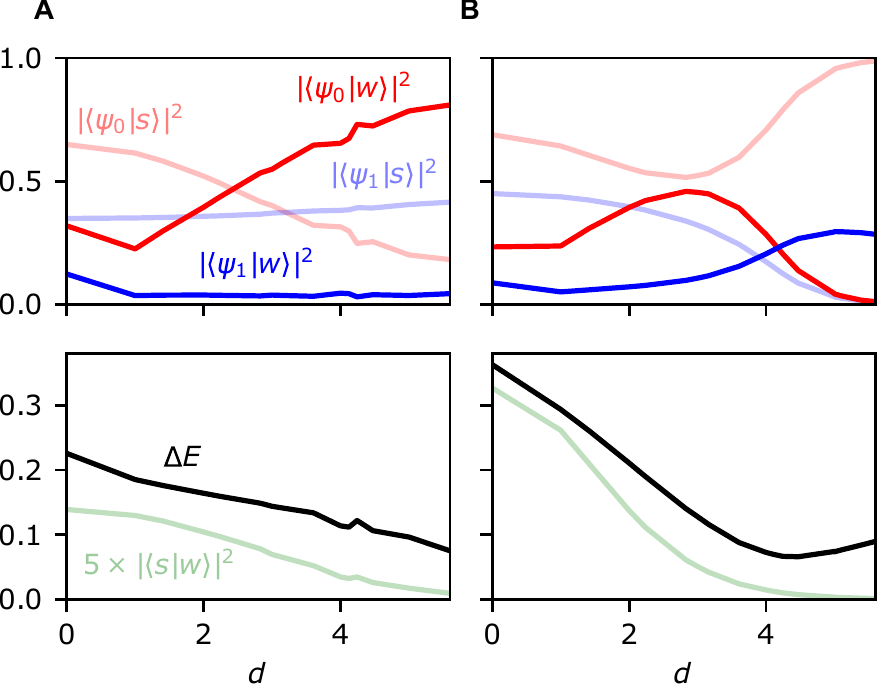}
    \caption{Level structure for different oracle positions. a) Closed boundaries, as in Fig.~\ref{sfig:bands}d. b) With a Gaussian confining potential, as in Fig.~\ref{sfig:bands}e. When translational symmetry is broken, the spectrum of the search Hamiltonian is dependent on the position of the oracle. Here, we plot the overlap between $\ket{\psi_{0,1}}$ and $\ket{s,w}$ (top), and the energy gap $\Delta E$ between states $\ket{\psi_{0,1}}$ (bottom), as a function of the magnitude of the oracle's distance from the lattice center, $ d $, in units of the lattice spacing $ a $. We find that these quantities are primarily related to the spatial structure in $\ket{s}$, and the resulting position-dependent overlap with $\ket{w}$ (bottom, $\times5$ scale).
    }
    \label{sfig:dists}
\end{figure}

In our experiment we must additionally consider the modified boundary conditions in our lattice. In the case of closed boundary conditions, or in the case of an additional confining potential, translational symmetry is broken, and the eigenstates of the lattice have distinct spatial structure. This can result in eigenstates that do not overlap with specific oracles, and thus are not affected by variations in $\gamma$ (Fig.~\ref{sfig:bands}a). In this case the search algorithm can still be interpreted in a similar way as in the case of cyclic boundary conditions, with the formation of a subspace composed of $\ket{\psi_0}$ and $\ket{\psi_1}$ supported by $\ket{s}$ and $\ket{w}$, but $\ket{\psi_0}$ and $\ket{\psi_1}$ need not be the ground and first excited states of the system. Furthermore, spatial structure in the ground state can also qualitatively modify the behavior of the search algorithm. For example, in the case of a finite 2D square lattice on a plane (Fig.~\ref{sfig:bands}d), $\ket{s}$ has a sine-profile along each axis with reduced overlap with nodes near the edges of the graph, since the edge nodes have reduced degree. This results in different behaviors for different oracle positions. For consistency, in this scenario we define a single critical value, $\gamma_c$, corresponding to the value of $\gamma$ that numerically minimizes $\Delta E$ for a centered oracle. For an $N=11\times11$ square lattice with closed boundary conditions, $\gamma_c=0.412$, which is similar to the critical value of $\gamma=0.430$ in the same lattice with cyclic boundary conditions. Studying the behavior of the closed system at $\gamma=\gamma_c=0.412$ as a function of oracle distance (Fig.~\ref{sfig:dists}a), we find that reduced overlap between $\ket{s}$ and $\ket{w}$ for oracles near the boundary results in reduced (increased) overlap between $\ket{\psi_0}$ and $\ket{s}$ ($\ket{w}$), and reduced $\Delta E$. Interestingly, the contributions from $\ket{s}$ and $\ket{w}$ to $\ket{\psi_1}$ remain relatively unchanged as a function of oracle distance.

Given this analysis, the main consequence of modified boundary conditions is how those conditions change the ground state $\ket{s}$. In this work, the confinement tweezer results in an approximately Gaussian ground state, which is similar to the sine-shaped ground state resulting from closed boundary conditions. Consequently, the behavior of the system under these conditions (Figs.\ref{sfig:bands}e, \ref{sfig:dists}b) is similar to those in the closed-boundary case, leading to a numerically computed critical value of $\gamma_c = 0.404 $, and thus an optimal oracle depth of $V_w = J_0/\gamma_c = 2.48J_0$. Note that in this case as we move away from the center of $\ket{s}$, the optimal depth of the applied oracle is reduced, which can be interpreted as the atoms effectively taking more time to reach the marked site. As a result, at intermediate oracle distances, and at this fixed value of $\gamma=\gamma_c$, the application of a deeper-than-optimal oracle leads to atoms being more tightly localized on the marked site in the state $\ket{\psi_0}$. This leads to the peak observed in Fig.~\ref{sfig:dists}b. At even further distances the oracle has very little overlap with $\ket{s}$, and effectively leaves the system unperturbed such that $\ket{\psi_0} \simeq \ket{s}$.

In the experiment, we operate at a constant, deeper-than-optimal oracle depth of $V_w = 12.55(65)J_0$. At these depths, even when the applied oracle fluctuates low due to the tweezer-lattice alignment concerns described in section~\ref{sec:alignment}, we expect the resultant dynamics to closely follow a sinusoidal oscillation of fixed amplitude and frequency. This means that a clear signal can be observed even when averaging over different oracle depths. By contrast, if operating closer to the optimal oracle depth, fluctuations to lower depths can lead to more complicated dynamics that are not strictly sinusoidal, and so averaging over these dynamics can lead to misleading signals. While the exemplary search data appearing in Fig.~\ref{sfig:images}fg is taken at the closer-to-optimal oracle depth of $3.1J_0$, and shows higher transfer into the marked site, in our apparatus, it is challenging to obtain sufficiently consistent data in this regime to properly characterize the coherent dynamics contributing to spatial search.

To calibrate the depth of the oracle, we operate at a fixed evolution time after the quench that is optimal based on calculations of the desired oracle depth, and perform the search procedure as a function of the applied oracle depth at a fixed location. At the depth that maximizes population on this marked site, we characterize the behavior of the quench as a function of both oracle position and time to confirm that the desired coherent behavior is occurring (see Fig.~\ref{fig:quench}b of main text). The resultant calibration of the oracle depth is in reasonable agreement with independently characterized values of the tweezer shapes and depths, and time evolution under these conditions matches up well with experimental observations (see Fig.~\ref{fig:quench}b of main text). Interestingly, the amplitude of both the calculated and observed oscillations are higher than what is suggested by the overlap between $\ket{\psi_1}$ and $\ket{w}$ alone, indicating that higher-energy eigenstates also play a role in the evolution.

While the position-dependent behavior of the system with respect to different oracles with fixed parameters seems antithetical to search, in practice, these changes are small enough that this search procedure still works for the system sizes explored in this work. However, these are important theoretical considerations for future works that attempt to operate in a regime with an asymptotic speedup, since different boundary conditions might change the asymptotic scaling of the algorithm.

\section{Prospects for multi-particle quantum walks and genuine quantum speedups}

A continuous-time quantum walk of a single particle on any graph with $N$ nodes is fully captured by a classical wave equation involving $N$ coupled oscillators, including the dynamics resulting in the $O(\sqrt{N})$ runtime for search in a sufficiently connected graph~\cite{grover_coupled_2002}. As such, to understand when a genuine asymptotic quantum speedup occurs, it is important to consider the physical resources required by a given algorithm in addition to its runtime.

In the standard oracular model of spatial search~\cite{aaronson_quantum_2003-1}, an algorithm must alternate between queries that determine whether the current location is marked, and operations that can only perform local interactions on the graph. Schematically, one can think of this as the case where a single local ``robot'' is allowed to traverse a memory that is spread out in physical space, and report the result of the computation back to the experimenter. In this model, any classical algorithm must make $O(N)$ queries to find the marked item, even if the graph is complete. The classical wave dynamics that reproduce the behavior of a single-particle quantum walk can be classically simulated in this model only with significant overhead. As a result, in this setting there can be an asymptotic quantum speedup for search by quantum walk even when working with a single particle (or robot) traversing a graph whose nodes correspond to physical locations in space.

A distinct formulation of the search problem is to find one of ${N=2^n}$ items using a single register of $n$ bits. In the classical case, one can check only one item at a time, leading to a runtime of $O(N)$. The $O(\sqrt{N})$ runtime of Grover's search on $N$ items using a register of $n$ qubits leads to the oft-quoted quadratic speedup. However, unlike in the case of the quantum robot, in this setting the speedup is dependent on the exponential scaling of Hilbert space with additional qubits. In this work each state in Hilbert space corresponds to a physical location in a real lattice, and so one could argue that the physical resource cost scales like $O(N)$ instead of $O(n)$. Additionally, we read out the position of the walker by performing a parallel measurement on all $ N $ sites in the lattice. In the classical case, this would correspond to having access to a register with $O(N)$ bits that can be read out in parallel. In this setting, one can simply assign a bit to each item in the search space, flip the bit corresponding to the marked item, and read out the whole register to find the marked item in a runtime of $O(1)$. This illustrates that, in some settings, the speedup associated with quantum search algorithms requires both reduced runtime, and a state space that scales rapidly with physical resources.

As stated in the main text, one way of achieving this scaling is to extend this work to multiple particles, where for $n$ particles and $m$ lattice sites, the state space has dimension $N=m^n$ for distinguishable particles, and $N= \binom{n+m-1}{m-1}$ for indistinguishable bosons. This approach also solves the issue of limited connectivity, since if these particles occupy a 2D square lattice, the full system maps to a square lattice graph with dimension $2n$. In this case the statistics of the particles change the size and boundary of the graph, but not its connectivity, suggesting that with the ability to engineer an oracle that marks the appropriate multi-particle state, a runtime of $O(\sqrt{N})$ could be achieved with just three or more indistinguishable bosons in a 2D lattice.

The bosonic $\rm^{88}Sr$ atoms used in this work are ideally suited to implementing such experiments due to their vanishingly small scattering cross section in the ground state~\cite{martinez_de_escobar_two-photon_2008}. The cooling performance and tweezer-based control demonstrated here suggest that experiments involving tens to hundreds of non-interacting and indistinquishable bosons could be achieved in this platform while maintaining reasonable post-selection rates. The ability to modify the sites that these bosons occupy in real time makes this an attractive platform for studying quantum optics problems of interest, including programmable boson sampling~\cite{aaronson_computational_2010}. This could be extended to more lattice sites and particles through further optimization of the lattice potential, and the adoption of more advanced cooling techniques~\cite{kaufman_cooling_2012, brown_hyperpolarizability_2017-1}.

Combining multi-particle tunneling with the tunable, non-local interactions provided by Rydberg dressing~\cite{guardado-sanchez_quench_2021} provides one route towards implementing the appropriate oracle for search by multi-particle quantum walks. For example, local Rydberg dressing beams could be used to mark a subset of sites in the lattice, such that interactions between atoms on these marked sites shifts the desired multi-particle state. Such interactions could also natively implement search oracles that effectively run a subroutine that checks the verifier to NP-complete problems of interest while maintaining a polynomial quantum speedup~\cite{anikeeva_number_2021-1}. In the context of quantum simulation, this combination of tunable interactions and itinerance could also enable the study of a broad class of extended Hubbard models which are thought to capture a range of exotic behaviors including supersolidity, and unconventional superconductivity involving charge fluctuations~\cite{onari_phase_2004, ohgoe_ground-state_2012}.

When working with multiple particles, control of the optical clock degree of freedom would also act as a powerful tool for tuning distinguishability and changing the behavior of multi-particle interference. For example, the recently demonstrated capability to engineer clock-entangled states via Rydberg interactions in this platform~\cite{schine_long-lived_2021} offers a route towards tuning the statistics associated with these atoms. In this case, the symmetry of the prepared entangled state enforces a symmetry on the tunneling, allowing (bosonic) $\rm^{88}Sr$ atoms to simulate fermionic interference, as well as certain classes of anyonic problems~\cite{wilczek_magnetic_1982, wilczek_quantum_1982, sansoni_two-particle_2012, matthews_observing_2013}. The vastly different scattering cross sections associated with ground- and clock-state $\rm^{88}Sr$ atoms~\cite{lisdat_collisional_2009, traverso_inelastic_2009} could also offer a route towards simulating hard core bosons~\cite{melko_supersolid_2005, wessel_supersolid_2005}, or dual-species Hubbard models~\cite{kuklov_counterflow_2003}. Incorporating a long-lived internal degree of freedom such as this one is of further interest in quantum networking, where tunneling and interference can be used to transmit a qubit through a network with imperfect switching~\cite{chakraborty_spatial_2016}.

\end{document}